\documentclass[10pt]{article}

\usepackage{amssymb,amsfonts}
\usepackage[mathscr]{eucal}
\usepackage{amsmath}
\usepackage[dvips]{graphicx}
\allowdisplaybreaks[1]

\newtheorem{assumption}{Assumption}

\newtheorem{theorem}{Theorem}[section]
\newtheorem{lemma}{Lemma}
\newtheorem{remark}{Remark}
\newtheorem{example}{Example}

\newenvironment{proof}{{\it Proof. }}{{ 
 \hfill{$\square$}}  \vspace{0.0cm} \medskip}

\newcommand{\DP}{DP}

\newcommand{\cl}{{\mathrm h}}

\newcommand{\orth}{\perp}

\newcommand{\id}{{ \rm id}}

\newcommand{\bN}{{\mathbb N}}
\newcommand{\bR}{{\mathbb R}}
\newcommand{\bC}{{\mathbb C}}
\newcommand{\bZ}{{\mathbb Z}}

\newcommand{\cA}{{\mathcal A}}
\newcommand{\cS}{{\mathscr S}}

\newcommand{\cR}{{\mathcal R}}
\newcommand{\cD}{{\mathcal D}}

\newcommand{\ct}{{\mathfrak t}}

\newcommand{\cG}{{\mathfrak g}}

\newcommand{\CW}{{\mathcal C}}

\newcommand{\cg}{c_{\cG}}

\newcommand{\ck}{{\ct^{\orth}}}
\newcommand{\cP}{{\mathcal P}}

\newcommand{\eps}{\epsilon}

\newcommand{\cW}{{\mathcal W}}

\DeclareMathOperator{\sing}{sing}

\DeclareMathOperator{\shift}{shift}

\DeclareMathOperator{\Isom}{Isom}

\DeclareMathOperator{\Image}{Image}
\DeclareMathOperator{\arc}{arc}

 \DeclareMathOperator{\Mat}{Mat}

\DeclareMathOperator{\Ad}{Ad}
\DeclareMathOperator{\ad}{ad}

\DeclareMathOperator{\aff}{aff}

 \DeclareMathOperator{\WLO}{WLO}

\DeclareMathOperator{\wind}{wind}

\DeclareMathOperator{\sgn}{sgn}
\DeclareMathOperator{\Diff}{Diff}
\DeclareMathOperator{\rank}{rank}

\DeclareMathOperator{\gleam}{gl}

\usepackage[dvips]{graphicx}

\def\a{\alpha}
\def\b{\beta}
\def\d{\delta}

\def\g{\gamma}

\def\vf{\varphi}

\def\l{\lambda}

\def\m{\mu}
\def\n{\nu}
\def\s{\sigma}
\def\S{\Sigma}

\def\r{\rho}

\def\th{\theta}

\newcommand{\sm}[1]{\mbox{\scriptsize #1}}

\renewcommand{\@}[1]{\sqrt{#1}}
\newcommand{\Tr}{{\mbox{Tr}}}
\newcommand{\be}{\begin{equation}}
\newcommand{\ee}{\end{equation}}
\newcommand{\bea}{\begin{eqnarray}}
\newcommand{\eea}{\end{eqnarray}}
\newcommand{\nn}{\nonumber}

\newcommand{\eq}[1]{(\ref{#1})}
\def\nn{\nonumber\\}

\def\ffract#1#2{\raise .35 em\hbox{$\scriptstyle#1$}\kern-.25em/
\kern-.2em\lower .22 em \hbox{$\scriptstyle#2$}}

\def\half{{1\over2}\,}

\begin{document}

\title{Chern-Simons theory and the quantum Racah formula}

\author{Sebastian de Haro \quad and \quad Atle Hahn}


\maketitle

\begin{abstract}
We generalize several  results  on Chern-Simons models on $\Sigma
\times S^1$ in the so-called ``torus gauge'' which were obtained
 in  \cite{Ha4} (= arXiv:math-ph/0507040) to the case of
general (simply-connected simple compact) structure groups and
general link colorings. In particular,  we give a non-perturbative
evaluation of the  Wilson loop observables corresponding to  a
special class of simple but non-trivial links and show that their
values  are given by Turaev's shadow invariant.
  As a byproduct we obtain a heuristic path integral derivation of
 the quantum Racah formula.
\end{abstract}

\section{Introduction}
\label{intro}

 In 1988 E. Witten  succeeded in defining, on a physical level of rigor,
  a large class of new 3-manifold and link invariants
  with the help of the heuristic Chern-Simons path integral,   cf.  \cite{Wi}.  Later   a rigorous definition of
these invariants was given, cf.  \cite{ReTu1,ReTu2} and part I of
\cite{turaevbook}.
 The approach in \cite{ReTu1,ReTu2} is based on the representation
theory of quantum groups\footnote{in fact, the approach in
\cite{ReTu1,ReTu2,turaevbook} is  more general, cf. Remark
\ref{rm_mod_cat} below}
 and uses surgery techniques on the base manifold. A related approach is
the so-called ``shadow world'' approach   (cf. \cite{ReKi,TuVi,Tu2}
and part II of \cite{turaevbook}), which also works with quantum
groups  but replaces  the use of surgery operations
 by certain combinatorial arguments leading to finite ``state
sums''.\par
 It is an  open problem (cf., e.g.,  p. 2 in \cite{Freed} and Problem (P1) in \cite{Ha4})
  how the  rigorous approaches using quantum groups are related to Witten's path integral
approach. This problem  is interesting by itself.
Moreover, one can expect that the solution
of this problem will lead to some progress towards
the solution of one of the central open problems in the field,
namely the question if/how one can make rigorous sense of the path
integral expressions used in the heuristic treatment in \cite{Wi}
(cf. Sec. \ref{sec6} below for additional comments).\par

The results in \cite{Ha4}, which were obtained by extending the
work in \cite{BlTh1,BlTh2,BlTh3,Ha3b} in a suitable way, suggest
that the key for establishing a direct relationship between the CS
path integral and the two quantum group approaches mentioned above
 is the so-called ``torus gauge fixing'' procedure,
introduced in \cite{BlTh1} for the study of CS models on base
manifolds $M$ of the form $M=\Sigma \times S^1$. Indeed, already
in \cite{BlTh1} it was demonstrated that in the torus gauge
setting the evaluation of the Wilson loop observables (WLOs) of
special links consisting exclusively of ``vertical loops''
naturally leads to the S-matrix expressions on the right-hand side
of the so-called Verlinde formula, cf. expression \eqref{rhs_fusion_rules}
below and Remark \ref{rm_BlTh} in Sec. \ref{sec7}. In \cite{Ha4} it
was then shown how to treat the case of general links within (a
suitably modified version of) the torus gauge setting. Moreover,
it was shown that in the special case $G=SU(2)$
 the evaluation of
the Wilson loop observables of loops without double points
naturally leads to  the gleam factors and the summation over
(admissible) ``area colorings'' present in Turaev's formula for
the shadow invariant (cf. Eq. \eqref{eq_shadow_SU2} below). In the
present paper we will generalize the results in \cite{Ha4} to
general (simply-connected simple compact) groups $G$ and to links
with arbitrary ``colors'', i.e. equipped with arbitrary
representations (and not only the fundamental representation as in
\cite{Ha4}). As a result we will be able to demonstrate that
within the torus gauge setting also the fusion coefficients (i.e. the numbers $N_{jl}^i$ in Eq.
\eqref{eq_XL3})  in
Turaev's formula  for the shadow invariant appear naturally when
links without double points are studied.\par

We mention here that Turaev's shadow invariant also appears in
the evaluation of a purely two-dimensional quantum field theory,
 namely $q$-deformed Yang-Mills theory on a Riemannian surface $\S$ \cite{deHa1}.
  The connection of the latter with Chern-Simons on $S^1$-bundles over $\S$,
  of which $S^1\times\S$ is a special case, was developed in \cite{deHa2,deHa1,AOSV,deHa3,deHaTi,BlTh4}.
  The algebraic lattice formulation of $q$-deformed two-dimensional Yang-Mills has been worked out for real $q$
  and not for $q$ being a root of unity \cite{BR}.
  Although we will not further develop the connection to this two-dimensional theory in this paper,
  we note that the intermediate expressions we obtain in our evaluation of the Chern-Simons path integral
   are those of $q$-deformed two-dimensional Yang-Mills.
  In turn, the path integral formulation of the simpler two-dimensional quantum field theory may be helpful
  in defining the Chern-Simons path integral on non-trivial bundles over $\S$ \cite{BlTh4,BW}.

The paper is organized as follows. In Subsec. \ref{subsec2.1} we
first recall some important concepts and construction from Lie
theory. In Subsec. \ref{subsec2.2} we then introduce
 some concepts from Conformal Field Theory
and the theory of affine Lie algebras  which played a role in \cite{Wi}.
  In Sec. \ref{sec3} we reformulate
Turaev's shadow invariant for manifolds of the form $\Sigma \times
S^1$ using the notation from Sec. \ref{sec2}.
  In Secs. \ref{subsec4.1}--\ref{subsec4.3} we  recall some of the
results obtained in \cite{BlTh1,BlTh2,BlTh3,Ha3b,Ha3c,Ha4} on
Chern-Simons models on $\Sigma \times S^1$ in the ``torus gauge''
and  in Subsec. \ref{subsec4.4} we then generalize the
calculations in \cite{Ha4} for the WLOs for links without double
points to the case of general (simple simply-connected compact)
groups $G$ and arbitrary link colorings.
 In Sec. \ref{sec5}  we   show that the finite state
sums appearing in Sec. \ref{sec4} are equivalent to the state sums
in the shadow invariant. In other words, the values of the WLOs
obtained in Sec. \ref{sec4} agree exactly with the values obtained
by applying the shadow invariant to the corresponding links, cf.
Eq. \eqref{eq_WLO=shadowinv}.
In Sec. \ref{sec7} we then show that
by reversing the order of arguments used in Secs. \ref{sec4} and
\ref{sec5} one can obtain a path integral derivation of the
so-called quantum Racah formula (cf. Eq. \eqref{eq_quantum_racah}
below). We conclude the paper with a brief
outlook in Sec. \ref{sec6}.

\section{Algebraic preliminaries}
\label{sec2}

\subsection{Concepts from classical Lie theory}
\label{subsec2.1} Let $G$ be a  simply-connected and simple
compact Lie group and $\cG$ its Lie algebra. Moreover, let $T$ be
a maximal torus of $G$ and  $\ct$ the Lie algebra of $T$. (We will
keep $G$ and $T$ fixed for the rest of this paper).

\begin{itemize}
\item $(\cdot,\cdot)$ denotes the Killing metric  on $\cG$
  normalized such that -- after making the identification
  $\ct$ and $\ct^*$  with the help of  $(\cdot,\cdot)$ --
  we have   $(\alpha,\alpha)=2$ if $\alpha$ is a long real root.

\item We set $r:=\dim(\ct)=\rank(\cG)$. $\pi_{\ct}:\cG \to
\ct$ will denote the $( \cdot, \cdot )$-orthogonal projection and
$\ck$  the $(\cdot,\cdot)$-orthogonal complement of $\ct$ in
$\cG$.

\item $\cR \subset \ct^*$ will denote the set of real roots associated to ($\cG, \ct)$
 and $\Check{\cR}$ the set of  real coroots, i.e.  $\Check{\cR}$
 is given by  $\Check{\cR} := \{\Check{\alpha} \mid \alpha \in \cR\} \subset \ct$
       where $\Check{\alpha}: = \frac{ 2 \alpha}{( \alpha,  \alpha)}$.
 Let $\Lambda \subset \ct^*$ denote the real weight lattice associated to $(\cG,\ct)$,
 i.e. $\Lambda$ is given by
\begin{equation}
\Lambda:= \{\lambda \in \ct^* \mid \lambda(\Check{\alpha}) \in \bZ \text{ for all } \alpha \in \cR\}
\end{equation}
$\Lambda_{\Check{\cR}} \subset \ct$ will denote the lattice generated by the real coroots.

 \item A Weyl chamber is a connected component of $\ct \backslash \bigcup_{\alpha \in \cR} H_{\alpha}$
 where  $H_{\alpha}:= \alpha^{-1}(0)$.
 A Weyl alcove (or ``affine Weyl chamber'') is a connected component
 of the set\footnote{note that in \cite{Ha3b} we used
the notation $\ct'_{reg}$ instead of $\ct_{reg}$.}
 $\ct_{reg}:=  \ct \backslash \bigcup_{\alpha \in \cR, k \in \bZ} H_{\alpha,k}$
 where $H_{\alpha,k}:= \alpha^{-1}(k)$.

 \item Let $\cW$  denote the Weyl group (associated to $\cG$ and $\ct$), i.e.
  the group of isometries of $\ct \cong \ct^*$
   generated by the orthogonal reflections on the hyperplanes $H_{\alpha}$, $\alpha \in \cR$, defined
   above.
 $\cW_{\aff}$ will denote the affine Weyl group, i.e.
  the group of isometries of $\ct \cong \ct^*$
   generated by the orthogonal reflections on the hyperplanes $H_{\alpha,k}$, $\alpha \in \cR$, $k \in \bZ$,
    defined  above\footnote{Equivalently, one can define $\cW_{\aff}$ as the group of isometries of $\ct \cong \ct^*$
generated by
    $\cW$ and the translations associated to the coroot lattice $\Lambda_{\Check{\cR}}$}.
 For $\tau \in \cW_{\aff}$ we will denote the sign
  of $\tau$ by  $\sgn(\tau)$.

\end{itemize}

\noindent In the sequel let us fix a Weyl chamber $\CW$. Let $P$
denote the unique  Weyl alcove which is contained in $\CW$ and has
$0 \in \ct$ on its boundary.

\begin{itemize}
\item Let $\cR_+$ denote the set of positive roots, i.e.  $\cR_+ := \{ \alpha \in \cR \mid (\alpha,x) \ge 0 \text{ for all }
 x \in \overline{\CW}\}$,
 and let $\Lambda_+$ denote the set of ``dominant  weights'', i.e.
     $\Lambda_+ := \Lambda \cap \overline{\CW}$.

\item For $\lambda \in  \Lambda_+$ let  $\rho_{\lambda}$   denote
the (up to equivalence) unique
 irreducible complex representation of $G$ with highest weight $\lambda$ and
$\chi_{\lambda}$  the  character
     corresponding to $\rho_{\lambda}$.
The  multiplicity   of the global weight associated to $\mu$
   in $\chi_{\lambda}$ will be denoted by $m_{\lambda}(\mu)$, i.e.
 we have
 \begin{equation} \label{eq_multiplicities}  \chi_{\lambda}(\exp(b)) =
\sum_{\mu \in \Lambda} m_{\lambda}(\mu) e^{2\pi i (\mu,b)}\text{
\quad for all $b \in \ct$}
\end{equation}

\item   $\rho$ will denote the half-sum of the positive roots
 and $\theta$  the unique  long root in the Weyl chamber $\CW$.
 The dual Coxeter number $\cg$ of $\cG$ is given by\footnote{
 note that $\cg= 1 + (\theta,\rho) =  \tfrac{1}{2} (\theta,\theta + 2\rho) =
  \tfrac{1}{2} C_2(\theta)$.
 If we had normalized
 the Killing form $(\cdot,\cdot)$ such that the long roots have length $1$  we would
  have $\cg=  C_2(\theta)$, i.e. $\cg$ would then be the Casimir element associated to
  the adjoint representation.
  }
 \be
 \cg= 1 + (\theta,\rho)
 \ee

\item For each $\lambda \in \Lambda_+$ we set
\be C_2(\l):=(\l,\l+2\r)
\ee
i.e., $C_2(\lambda)$  is the
second Casimir element (w.r.t. to  the inner product $(\cdot,\cdot)$)
 corresponding to the irreducible representation of $\cG$ with highest weight $\lambda$.

\item For $\lambda \in  \Lambda_+$ let $\overline{\lambda} \in
\Lambda_+$ denote the weight conjugated to $\lambda$ and
$\lambda^* \in \Lambda_+ $ the weight conjugated to $\lambda$
``after applying a shift by $\rho$''. More precisely, $\lambda^*$
is given by $\lambda^* + \rho = \overline{\lambda + \rho}$.
\end{itemize}

\begin{remark} \label{rem1} \rm
Let $I \subset \ct$  denote the ``integral lattice'', i.e.
 $I:= \ker(\exp_{| \ct})$.
 From the assumption that  $G$ is  simply-connected
it follows that $I$ coincides with the lattice $\Lambda_{\Check{\cR}}$ generated by the
    real coroots  so
    the weight lattice $\Lambda$ associated to $(\cG,\ct)$  coincides
    with the weight lattice $I^*$ of $(G,T)$ given by
 $I^*:= \{\alpha \in \ct^* \mid \alpha(x) \in \bZ \text{ for all } x \in I\}$.
\end{remark}

\subsection{Some concepts from CFT, the theory of affine Lie algebras, and the theory of
quantum groups}
\label{subsec2.2}

Let us fix $k \in \bN$  (the ``level'').
\begin{itemize}
\item  We set
\begin{equation}  \Lambda_+^k :=  \{ \l \in \Lambda_+  \mid  (\l,\th)\leq k \}
\end{equation}

\item Let $\Isom(\ct)$ denote the group of isometries of the
Euclidean vector space $(\ct,(\cdot,\cdot))$ and let $i:\Isom(\ct)
\to \Isom(\ct)$ denote  the automorphism of $\Isom(\ct)$ given by
\begin{equation} \label{eq_Wq_Waff}
i(\tau)(b) = (k+ \cg) \cdot \tau((b+\rho)/(k + \cg)) - \rho
\end{equation}
for all $b \in \ct$ and $\tau \in \Isom(\ct)$.
 We set\footnote{$\cW_k$ coincides with the subgroup of $\Isom(\ct)$ which is
generated by the orthogonal reflections on the $\rho$-shifted
   hyperplanes $H_{\alpha}-\rho$, $\alpha \in \cR_+$, and the hyperplane
   $\{y \in \ct \mid (y,\theta)=k+\cg\} - \rho = \{x \in \ct \mid (x,\theta)=k+1\}$,
   thus $\cW_k$ is the same as the group $\cW_0$ in
   \cite{Sawin99}.}
\begin{equation} \label{eq_def_Wq} \cW_k := i(\cW_{\aff}) \subset \Isom(\ct)
\end{equation}
(the ``($\rho$-shifted) quantum Weyl group corresponding to the level
$k$'') and
$$\sgn(\tau):=\sgn(i^{-1}(\tau)) \quad  \text{ for } \tau \in \cW_k$$

\item Let $C$, $S$, and  $T$   be the $\Lambda_+^k \times \Lambda_+^k$ matrices with complex entries
   given by
   \begin{subequations}
\begin{align} 
\label{eq_C}  C_{\l\m} & := \delta_{\l \m^*}, \\
\label{eq_T} T_{\l\m}  & := \delta_{\l\m}\,e^{\pi i C_2(\lambda )\over k+\cg}
\cdot e^{-{\pi i c\over 12}},\\
\label{eq_S} S_{\l\m} & :={i^{|\cR _{+}|}\over(k+\cg)^{r/2}}\,|\Lambda/ \Lambda_{\Check{\cR}}|^{-\half}
\sum_{w\in \cW} \sgn(w)e^{{2\pi i\over k+\cg}(\l + \rho ,w\cdot
(\m + \rho) )}
\end{align}
\end{subequations}
for all $\l, \m \in \Lambda_+^k$, cf. Eqs. (14.216), (14.217), and (14.229) in \cite{diF}
and compare also Sec. II.3.9 in \cite{turaevbook} where a slightly
different convention is used\footnote{the matrix $C$ is called $J$ in \cite{turaevbook}.
Moreover, the matrix $S$ in \cite{turaevbook}
differs from the matrix $S$ in Eq. \eqref{eq_S} by a multiplicative constant   ``$\cD$'', cf.
the ``Notes'' at the end of Chap. II in \cite{turaevbook}}.\par

 We remark that the  factor $e^{-{\pi i c\over 12}}$ with $c: = \dim(\cG) \cdot {k \over (k+\cg)}$
 appearing in Eq. \eqref{eq_T}  is not really essential for the present paper.
In particular, the definition of $|X_L|$ in Eq. \eqref{shadow} below and Theorem \ref{theorem} below (and also
 the computations in Sec. \ref{sec7} below) are  not affected\footnote{cf. step $(**)$ in the proof
of part iii) of Lemma \ref{lemma1} below} if we omit this factor.
The advantage of including the factor $e^{-{\pi i c\over 12}}$ in Eq. \eqref{eq_T}
is that Eq. \eqref{defST2} below holds in the ``strict sense''  and not only in the ``projective sense''.
This point  simplifies the computations in our examples in Sec. \ref{subsec3.2}.\par

One can  prove  (cf. Eqs. (10.206), (10.216),  (14.228) and Exercise 14.14 in \cite{diF} and\footnote{in Sec. II.3.9 in \cite{turaevbook}  the expression
  $ \cD^{-1} \triangle$ appears.
  The computations in \cite{turaevbook} imply Eq. \eqref{defST2}
  provided that    $ \cD^{-1} \triangle = e^{-{\pi i c\over 4}}= (e^{-{\pi i c\over 12}})^{3}$.
  In view of the results in \cite{Finkelberg} this is exactly
  what one expects} Sec. II.3.9 in \cite{turaevbook})  that
\begin{subequations} \label{defST_0}
 \begin{align}
  \label{defST}  S^2 &= C,\\
   \label{defST2} (ST)^3& = C
\end{align}
 \end{subequations}
  In particular, $S$
is invertible.\par

\item For $\lambda \in \Lambda^k_+$ we set \be \label{eq_def_dim}
\dim \lambda := \frac{S_{\lambda 0}}{S_{00}} \overset{(*)}{=}
\prod_{\a \in \cR_+}{\sin{\pi(\l+\r,\a)\over k +
\cg}\over\sin{\pi(\r,\a)\over k + \cg}} \ee

Here $(*)$ follows from $\frac{S_{\lambda 0}}{S_{00}} =
\frac{A(\rho)(\lambda + \rho)}{A(\rho)(\rho)}$  and the
relation\footnote{cf. part iii) of  Theorem 1.7 in Chap. VI of \cite{Br_tD}} $\delta(b) =A(\rho)(b) $ where
 \begin{align*} A(b')(b) & := \sum_{w\in \cW} \sgn(w) e^{2\pi i (b', w
\cdot b)}\\
\delta(b) & :=  \prod_{\beta \in \cR_+}  (e^{ \pi i \beta(b)}
-e^{- \pi i \beta(b)})=
 \prod_{\beta \in \cR_+}  2i \sin( \pi (\beta,b)).
 \end{align*}
for all $b , b' \in \ct$.

\item For $\lambda, \mu, \nu \in \Lambda^k_+$ we define the
``fusion coefficients'' $N_{\l\m\n}$ and $N_{\m\n}^\l$ by
 \be \label{rhs_fusion_rules}
N_{\l\m\n}:=\sum_{\s \in \Lambda_+^k} {S_{\l\s}
S_{\m\s}S_{\n\s}\over S_{0\s}} \ee and
  \be N_{\m\n}^\l:= N_{\l^*
\m\n}\ee
 Observe that  Eq. \eqref{defST} implies  $ N_{\m
0}^\n=\d_\m^\n$.

\end{itemize}

Let us  motivate the use of the term  ``fusion
coefficients'' above.  Let $\hat{\cG}$ denote the (non-twisted) affine Lie
algebra corresponding to $\cG_{\bC}:= \cG \otimes_{\bR} \bC$ (cf. Eq. (14.13) in \cite{diF}) and
let  $\hat{N}_{\l\m}^\n$ be the fusion coefficients of the modular tensor category
 based on the integrable representations of $\hat{\cG}$
  at level $k$.
Similarly, let $\Check{N}_{\l\m}^\n$ be the fusion coefficients in the modular tensor category
  constructed in \cite{And1,AndPar} using the representation theory of the quantum group $U_q(\cG_{\bC})$
  with\footnote{cf., e.g., \cite{Sawin99,Sawin05}.
  Sometimes in the literature a different convention is used
   for the definition of $U_q(\cG_{\bC})$,  which leads to the formula
    $q:=  e^{\frac{\pi i}{ D (k+\cg)}}$ where   $D$ is the quotient of the square lengths
of the long and the short roots of $\cG$ (cf., e.g.,
   the second page of the introduction in  \cite{Sawin03})}
$$q:=  e^{\frac{2 \pi i}{k+\cg}}$$
According to the famous ``Verlinde(-Moore-Seiberg) formula'' we have
\begin{subequations}
\begin{align} \label{eq_Verlinde}
\hat{N}_{\m\n}^\l & = N_{\m\n}^\l
 \end{align}
cf., e.g.,  \cite{Fuchs}. Similarly, according to the quantum group analogue
of the Verlinde formula (cf., e.g., Theorem 4.5.2 in Chap. II in \cite{turaevbook}) we have\footnote{of course
Eqs. \eqref{eq_Verlinde} and \eqref{eq_Verlinde2} imply  $\hat{N}_{\m\n}^\l  = \Check{N}_{\m\n}^\l $.
This is not surprising since the two modular tensor categories
mentioned above can be shown to be equivalent, cf.  \cite{Finkelberg}}
\begin{align}
  \label{eq_Verlinde2}
\Check{N}_{\m\n}^\l & = N_{\m\n}^\l
\end{align}
\end{subequations}
Moreover, in \cite{AndPar,Sawin03} it was proven that
  \begin{equation} \label{eq_quantum_racah0} \Check{N}_{ \gamma \alpha}^{\beta} = \sum_{\tau \in \cW_{k}}
\sgn(\tau) m_{\gamma}(\alpha-\tau(\beta))
\end{equation}
The last formula  can be considered
to be a ``quantum analogue'' of the classical Racah formula.
Following  \cite{Sawin03} we will call this formula  the ``(abstract) quantum Racah formula''
and the  formula
    \begin{equation} \label{eq_quantum_racah}
 N_{ \gamma \alpha}^{\beta} = \sum_{\tau \in \cW_{k}} \sgn(\tau) m_{\gamma}(\alpha-\tau(\beta)),
 \end{equation}
which follows from Eq.  \eqref{eq_quantum_racah0}  and \eqref{eq_Verlinde2}
will be called\footnote{Since for the derivation of
\eqref{eq_quantum_racah}
 we used both the Verlinde formula \eqref{eq_Verlinde2} and the (abstract) quantum-Racah formula
this name might be a little bit misleading.
 We could equally well call \eqref{eq_quantum_racah} the ``elementary  Verlinde formula''}
   the ``elementary quantum Racah formula''.

\section{The shadow invariant for links in $\Sigma \times S^1$}
\label{sec3}
\subsection{Definition}
\label{subsec3.1}

Let $\Sigma$ be an oriented surface, let  $L= (l_1, l_2,
\ldots, l_n)$, $n \in \bN$,  be a sufficiently regular  link in
 $\Sigma \times S^1$, and let $l^j_{S^1}$ resp.  $l^j_{\Sigma}$ denote the
projection of the loop $l_j$ onto the $S^1$-component resp.
$\Sigma$-component of the product $\Sigma \times S^1$.
  $L$ can be turned into a framed link
by picking for each loop $l_j$ the standard framing
described in Sec. 4 c) in  \cite{Tu2} (this framing was
 called ``vertical framing'' in \cite{Ha4}). We also assume that each loop $l_j$ is
colored with an element $\gamma_j$ of $\Lambda^k_+$.\par
 We  set $D(L):=(\DP(L),E(L))$ where $\DP(L)$ denotes the set of double points of
 $L$, i.e. the set of points $p \in \Sigma$ where the loops
 $l^j_{\Sigma}$, $j \le n$, cross themselves or each other,
 and $E(L)$ the set of
curves in $\Sigma$ into which the loops $l^1_{\Sigma},
l^2_{\Sigma}, \ldots, l^n_{\Sigma}$ are decomposed when being
``cut''  in the points of $\DP(L)$. Clearly, $D(L)$ can be
considered to be a finite (multi-)graph. We set $ \Sigma
\backslash D(L):=  \Sigma \backslash ( \bigcup_j
\arc(l^j_{\Sigma}))$.
 We assume that the set of connected
components of  $\Sigma \backslash D(L)$
 has only finitely many elements
$Y_0, Y_1, Y_2, \ldots, Y_{\mu}$, $\mu \in \bN$,  which we will
call the ``faces'' of  $\Sigma \backslash D(L)$.\\
As explained in \cite{Tu2} one can associate in a natural way a
 number $\gleam(Y_t)  \in \bZ$, called ``gleam'' of $Y_t$,
to each face $Y_t$ (for an explicit formula for the gleams in
 the special cases that will be relevant for us later see Eq. \eqref{eq_formel_gleam}
 below).
 We call  $X_L:= (D(L),
(\gleam(Y_t)_t)_{0 \le t \le \mu})$
  the ``shadow'' of $L$.\par
 Let $g \in E(L)$ be a fixed edge of the graph $D(L)$.
Note that, as each loop $l^j$ is oriented,  $g$ is an oriented
curve in $\Sigma$. On the other hand, as $\Sigma$ was assumed to
be oriented,
 each face $Y \in \{Y_0, Y_1, Y_2, \ldots, Y_{\mu}\}$ is an oriented surface
 and therefore also induces an orientation on its boundary
 $\partial Y$.\par
\begin{figure}[h]
\begin{center}
\includegraphics[height=1in,width=1.5in]{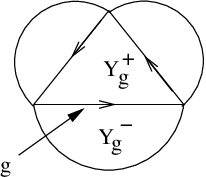}
\caption{} \label{trefoil}
\end{center}
\end{figure}

 There is a unique face $Y$,  denoted by
 $Y^+_g$ (resp. $Y^-_g$) in the sequel,
  such that $\arc(g) \subset \partial Y$ and,
  additionally,  the orientation on $\arc(g)$ described above
coincides with (resp. is opposite to) the orientation which is
obtained by restricting the orientation on $\partial Y$ to
$g$. In other words: $Y^+_g$ and $Y^-_g$ are the two\footnote{
note that if Assumption  \ref{assump2} below is not fulfilled then
possibly  $Y^+_g =  Y^-_g$, so in this case there is actually only
one such face}
 faces that ``touch'' the edge $g$, and  $Y^+_g$ (resp. $Y^-_g$)
is the face lying ``to the left'' (resp. ``to the right'') of $g$,
cf. Fig. \ref{trefoil}.

A mapping $\vf: \{Y_0, Y_1, Y_2, \ldots, Y_{\mu}\} \to
\Lambda^k_+$ will be called an  area coloring of $X_L$ (with
colors in $\Lambda^k_+$) and the set of all such area colorings
will be denoted by $col(X_L)$.
 We can now define the shadow invariant
$|\cdot|$ by\footnote{this coincides with the definition in
\cite{turaevbook} up to an overall normalization factor which will
be irrelevant for our purposes}
 \be\label{shadow}
|X_L|:= \sum_{\vf\in\sm{col}(X_L)}
|X_L|_1^{\vf}\,|X_L|_2^\vf\,|X_L|_3^\vf\,|X_L|_4^\vf~ \ee
  where
\begin{align} |X_L|_1^\vf&=\prod_Y(\dim(\vf(Y)))^{\chi(Y)}\\
|X_L|_2^\vf&= \prod_Y (v_{\vf(Y)})^{\gleam(Y)}\\
\label{eq_XL3}  |X_L|_3^\vf&= \prod_{g \in E(L)} N_{co(g)
\varphi(Y^+_g)}^{\varphi(Y^-_g)}.
  \end{align} where  $N_{jl}^i$
and $\dim(\cdot)$ are as in Subsec. \ref{subsec2.2}, where $co(g)$
denotes the color associated to the edge $g$ (i.e.
$co(g)=\gamma_i$ where $i \le n$ is given by $\arc(l^i_{\Sigma})
\supset g$) and where we have set $v_\l :=T_{\l\l}$ (here $T$ is,
of course, the $T$-matrix from Subsec. \ref{subsec2.2}).

\smallskip

  $|X_L|_4^\vf$ is defined in terms of quantum 6j-symbols associated to $U_q(\cG_{\bC})$,
cf. Chap. X, Sec. 1.2 and Chap. XI, Sec. 6.3 in \cite{turaevbook}. In  view of
Assumption \ref{assump1} below and the consequences that this
assumption has, cf. Eq. \eqref{eq_reduced_shadow} below, the
precise definition of $|X_L|_4^\vf$ in the general case will not be relevant in the
present paper.

\medskip

 For the rest of this paper, we will restrict ourselves to the
special situation where $L$ also fulfills the following two
assumptions.

\begin{assumption}  \label{assump1}
  The colored link $L$ has no double points,
 i.e. the projected loops
$l^1_{\Sigma},l^2_{\Sigma},\ldots,l^n_{\Sigma}$ are
non-intersecting Jordan loops in $\Sigma$.
\end{assumption}

\begin{assumption} \label{assump2}
Each $l^j_{\Sigma}$ is 0-homologous.
\end{assumption}

\noindent Assumptions \ref{assump1}  and \ref{assump2} have the
following consequences:

\begin{itemize}
\item For each $j \le n$ the set  $\Sigma \backslash
\arc(l^j_{\Sigma})$ has exactly two connected components. In the
sequel
 $R^+_j$  (resp. $R^-_j$)
will denote the connected component ``to the left'' (resp. ``to
the right'') of
 $l^j_{\Sigma}$, i.e.   $R^+_j$  (resp. $R^-_j$) is the unique
 connected component containing $Y^{+}_j$ (resp.   $Y^{-}_j$) where we have set
 \begin{equation}
 Y^{\pm}_{j} := Y^{\pm}_{l^j_{\Sigma}}
 \end{equation}
 (i.e. $Y^{\pm}_{j} =  Y^{\pm}_{g}$ where $g=l^j_{\Sigma}$).

\item  $\mu=n$, i.e.
 $\Sigma \backslash (
\bigcup_j \arc(l^j_{\Sigma}))$ has $n+1$ connected components
$Y_0, Y_1, \ldots, Y_{n}$

\item For each $Y \in \{Y_0, Y_1, Y_2, \ldots, Y_n\}$ we have
\begin{equation} \label{eq_formel_gleam}
\gleam(Y) = \sum_{j \text{ with } \arc(l^j_{\Sigma}) \subset
\partial Y}  \wind(l^j_{S^1}) \cdot \sgn(Y;l^j_{\Sigma})
\end{equation}
where $ \wind(l^j_{S^1})$ is the winding number of the loop
 $l^j_{S^1}$ and where $ \sgn(Y;l^j_{\Sigma})$
 is given by

 $$ \sgn(Y; l^{j}_{\Sigma}):=
\begin{cases} 1 & \text{ if  $Y \subset R^+_j$ }\\
-1 & \text{ if  $Y \subset R^-_j$ }\\
\end{cases} $$

\item According to the general definition of the shadow invariant
  in Chap. X, Sec. 1.2 in \cite{turaevbook}. Assumption
  \ref{assump1} implies $|X_L|_4^\vf =1$
   so   Eq. \eqref{shadow}
reduces to
 \begin{equation}
 \label{eq_reduced_shadow}
|X_L|=\sum_{\vf\in\sm{col}(X_L)} |X_L|_1^\vf |X_L|_2^\vf
|X_L|_3^\vf
\end{equation}

 \item ``Vertical'' framing for a loop $l_j$ in $\Sigma \times S^1$ (cf. the first paragraph
   of the present subsection) is equivalent to  what was called ``horizontal'' framing in Subsec. 5.2 in \cite{Ha4}
\end{itemize}

\begin{remark} \rm \label{rm_mod_cat}
\begin{enumerate}
\item The ``shadow invariant'' defined in \cite{turaevbook} is
more general than
 what we have defined here above.
 Our definition is the special case of Turaev's shadow
invariant where  the underlying modular tensor category is the one coming
from the representation theory
 of the quantum groups $U_q(\cG_{\bC})$, cf. Sec. \ref{subsec2.2}
 above.

\item In the special case $G=SU(2)$ one has $N^i_{jk} \in \{0,1\}$
  for all $i,j,k \in \Lambda^k_+$
  so $|X_L|_3^\vf \in \{0,1\}$ for each $\vf   \in col(X_L)$.
  Let us call $\vf \in
  col(X_L)$ ``admissible'' iff $|X_L|_3^\vf =1$
 and set $col_{adm}(X_L):= \{ \vf \in  col(X_L) \mid \vf
 \text{ admissible } \}$.
 Then we can rewrite Eq. \eqref{shadow} in the form
 \be \label{eq_shadow_SU2}
 |X_L|:= \sum_{\vf\in col_{adm}(X_L)}
|X_L|_1^{\vf}\,|X_L|_2^\vf\, |X_L|_4^\vf \ee
 If one compares this formula with Eqs. (5.7) and (5.8) in \cite{Tu2} (and the two equations before Theorem 6.1
  in \cite{Tu2})   it is easy to see that the ``shadow invariant'' that was defined
  in  \cite{Tu2} (and used in \cite{Ha4}) is
   the special case of the shadow invariant in the present
  paper which one obtains by taking $G=SU(2)$.
\end{enumerate}
\end{remark}

\subsection{Some examples}
\label{subsec3.2}

\begin{example} \label{Ex1} \rm
Let $\Sigma=S^2$ and  let  $L=((l_1,l_2,l_3), (\lambda,\mu,\nu))$
be a colored link in $\Sigma \times S^1$ such that
 $\wind(l^i_{S^1})=1$ for all
$i = 1,2,3$ and such that  the
 projection of $L$ onto
the surface $\Sigma$ looks like in the  following figure. Let, for
$i \in \{1,2,3\}$,
 $Y_i$ denote the face ``enclosed'' by $l_{\Sigma}^i$
 and let $Y_0$ denote the remaining face.
 Clearly, we have
 $\chi(Y_i)=1$ for $i \in \{1,2,3\}$ and $\chi(Y_0)=2-2g-3=-1$
 and
 $\gleam(Y_i)=1$ for $i \in \{1,2,3\}$ and $\gleam(Y_0)=-3$.
\begin{figure}[h]
\begin{center}
\includegraphics[height=1in,width=3in]{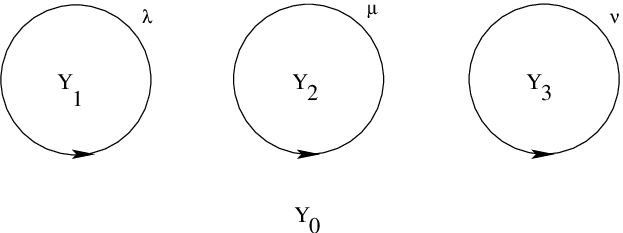}
\caption{} \label{3circles}
\end{center}
\end{figure}
So we obtain
 \bea
|X_{L}|&=&\sum_{\s_1\s_2\s_3\s_0 \in
\Lambda^k_+}\mbox{dim}(\s_1){\mbox{dim}}(\s_2){\mbox{dim}}(\s_3)(\mbox{dim}(\s_0))^{-1}\,N_{\s_1\l}^{\s_0}N_{\s_2\m}^{\s_0}N_{\s_3\n}^{\s_0}\,T_{\s_1\s_1}T_{\s_2\s_2}T_{\s_3\s_3}T_{\s_0\s_0}^{-3}\nn
&=&{T_{\l\l}T_{\m\m}T_{\n\n}\over T_{00}^3S_{00}^2}\,N_{\l\m\n}~.
\eea
 In deriving the last line, we used the following equation three
 times
\bea\label{fusioneqs} \sum_{\l \in \Lambda^k_+} \mbox{dim}(\l)\,
T_{\l\l}\,N_{\m\l}^\n &=&{1\over T_{00}S_{00}}\,(TST)_{\m\n} \eea
(Eq. \eqref{fusioneqs} follows from \eqref{defST_0} and
\eqref{rhs_fusion_rules}).
\end{example}

\begin{example} \label{Ex2}  \rm
Let again $\Sigma=S^2$ and  let  $L=((l_1,l_2,l_3),
(\lambda,\mu,\nu))$ be a colored link in $\Sigma \times S^1$ such
that
 $\wind(l^i_{S^1})=1$ for all
$i = 1,2,3$ and such that  the
 projection of $L$ onto
the surface $\Sigma$ looks like in Fig. \ref{smallcircles}.
 Then we have  $\chi(Y_1)=\chi(Y_3)=1$,  $\chi(Y_0)=\chi(Y_2)=0$
 and
  $\gleam(Y_1)=\gleam(Y_3)=1$,  $\gleam(Y_0)=-2$,   $\gleam(Y_2)=0$
where the faces $Y_0, Y_1, Y_2, Y_3$ are given as in Fig.
\ref{smallcircles}.
\begin{figure}[h]
\begin{center}
\includegraphics[height=1in,width=3in]{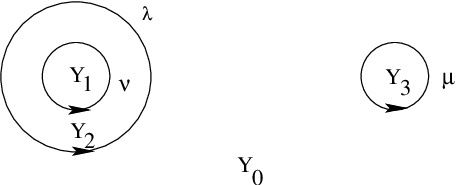}
\caption{} \label{smallcircles}
\end{center}
\end{figure}
 One obtains
  \be |X_{L}|=\sum_{\s_1\s_2\s_3\s_0 \in
\Lambda^k_+}\mbox{dim}(\s_1)\,\mbox{dim}(\s_3)\,N^{\s_2}_{\n\s_1}N^{\s_0}_{\l\s_2}
N^{\s_0}_{\m\s_3}\,T_{\s_1\s_1}T_{\s_3\s_3}T_{\s_0\s_0}^{-2}. \ee
The sums over $\s_1$ and $\s_3$ can be performed right away using
twice Eq.  \eq{fusioneqs}. We get
 \be \label{eq_1.step}
 |X_{L}| =
{T_{\m\m}T_{\n\n}\over
T_{00}^2S_{00}^2}\sum_{\s_2\s_0}T_{\s_2\s_2}T_{\s_0\s_0}^{-1}S_{\n\s_2}S_{\m\s_0}N_{\l\s_2}^{\s_0}~.
\ee
 Now  observe that
 \begin{multline} \label{eq_2.step}
\sum_{\s_2\s_0}T_{\s_2\s_2}T_{\s_0\s_0}^{-1}S_{\n\s_2}S_{\m\s_0}N_{\l\s_2}^{\s_0}
\overset{(*)}{=} {1 \over T_{\n\n}}
\sum_{\s_0\s}T_{\s_0\s_0}^{-1}T_{\s\s}^{-1}S_{\m\s_0}S_{\s_0\s}^{-1}
S_{\s\l}S_{\n\s}{1\over S_{\s0}}  \overset{(**)}{=} {T_{\m\m}
\over T_{\n\n}} N_{\l\m\n}
 \end{multline}
Here step $(*)$ follows from
 Eq. \eqref{rhs_fusion_rules} and
$STS=T^{-1}ST^{-1}$  (which in turn follows from Eq.
\eqref{defST_0})
 and  step $(**)$ follows from  Eq. \eqref{rhs_fusion_rules} and $ST^{-1}S^{-1}=TST$. From Eqs.
\eqref{eq_1.step} and \eqref{eq_2.step} we finally get \be
|X_{L}|={T_{\m\m}^2\over T_{00}^2S_{00}^2}\,N_{\l\m\n}~. \ee
\end{example}

The next example generalizes  the first two examples above.

\begin{example}  \label{Ex4} \rm
Let $\Sigma=S^2$ and  let  $L = (l_1,l_2, \ldots, l_{m+n+1})$ be a colored link in $\Sigma \times S^1$
consisting of $m+n+1$ loops with
colors $\nu_1,\nu_2,\cdots,\nu_m$, $\lambda$, and $\m_1,\m_2,\cdots,\m_n$ such that
 $\wind(l^i_{S^1})=1$ for all
$i = 1,2,\ldots, m+n+1$ and such that  the
 projection of $L$ onto
the surface $\Sigma$ looks like in the  following figure.
 \begin{figure}[h]
\begin{center}
\includegraphics[height=1.5in,width=3in]{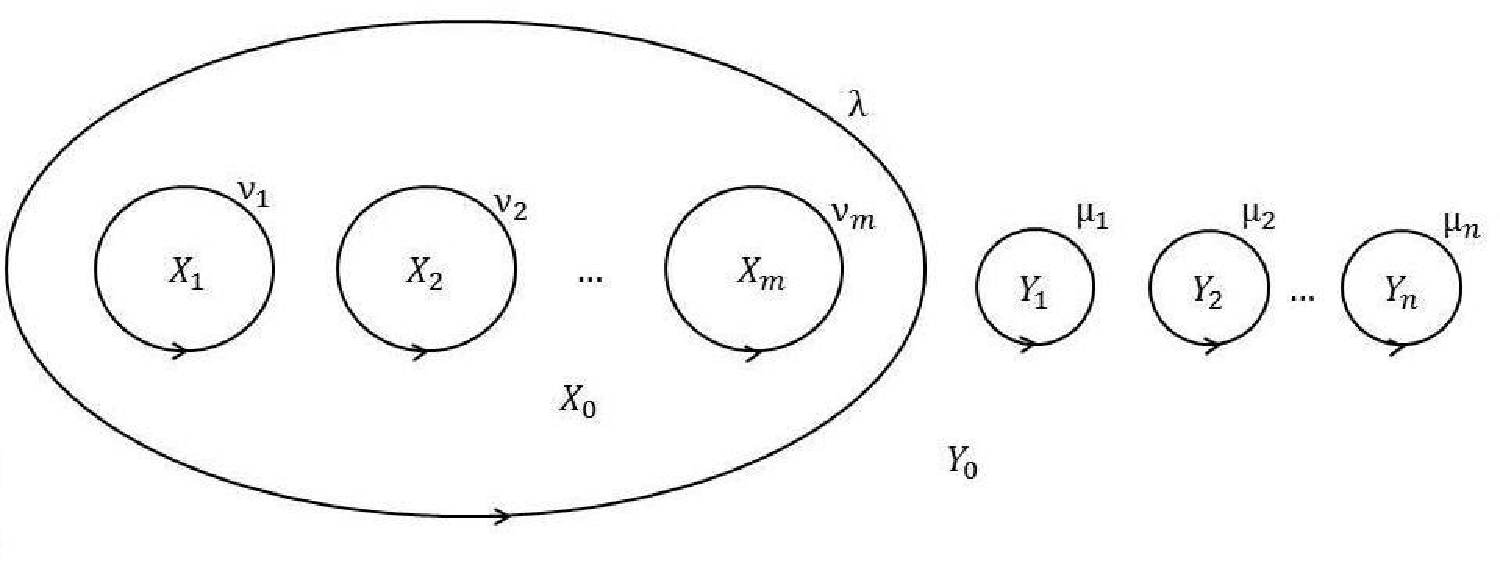}
\caption{}
\end{center}
\end{figure}

Let $X_1,X_2,\cdots,X_m$ be the faces encircled by
the first $m$ loops with colors  $\nu_1,\nu_2,\cdots,\nu_m$.
 Loop $l_{m+1}$ has color $\l$ and encircles the first $m$ loops.
  The face  ``inside'' this  loop (subtracting the (closure of the) faces of the $m$ loops from it) is $X_0$.
  ``Outside'' this group of loops  are $n$ more loops with colorings $\m_1,\m_2,\cdots,\m_n$
  encircling the faces $Y_1,Y_2,\cdots,Y_n$, respectively.

  \smallskip

The Euler characters are as follows:
\begin{subequations}
\begin{align}
\chi(Y_1)&=\chi(Y_2)=\cdots=\chi(Y_n)=\chi(X_1)=\chi(X_2)=\chi(X_m)=1\\
\chi(Y_0)&=1-n\\
\chi(X_0)&=1-m
\end{align}
\end{subequations}
and the gleams:
\begin{subequations}
\begin{align}
{\mbox{gl}}(X_1)&={\mbox{gl}}(X_2)=\cdots={\mbox{gl}}(X_m)={\mbox{gl}}(Y_1)={\mbox{gl}}(Y_2)=\cdots={\mbox{gl}}(Y_n)=1,\\
{\mbox{gl}}(Y_0)&=-n-1\\
{\mbox{gl}}(X_0)&=-m+1.
\end{align}
\end{subequations}
The value of $|X_L|$ is obtained by summing over all possible colorings of the faces. Using the
summation variables $\s_0,\s_1,\cdots,\s_n,\tau_0,\tau_1,\dots,\tau_m\in \Lambda_+^k$ we obtain
\begin{align}
|X_L|=\sum_{\s_0\cdots\s_n,\tau_0\cdots\tau_m}&{\mbox{dim}} \s_1 {\mbox{dim}}\s_2 \cdots\s_n {\mbox{dim}}\tau_1\cdots {\mbox{dim}}\tau_m ({\mbox{dim}}\s_0)^{1-n}({\mbox{dim}}\s_\tau)^{1-m}\times \nonumber \\
&\times N_{\s_1\m_1}^{\s_0}N_{\s_2\m_2}^{\s_0}
\cdots N_{\s_n\m_n}^{\s_0}N_{\s_1\m_1}^{\s_0}N_{\tau_1\nu_1}^{\tau_0}\cdots N_{\tau_m\nu_m}^{\tau_0}N_{\tau_0\l}^{\s_0} \nonumber \\
&\times T_{\s_1\s_1}T_{\s_2\s_2}\cdots T_{\s_n\s_n}T_{\s_0\s_0}^{-n-1}
T_{\tau_1\tau_1}T_{\tau_2\tau_2}\cdots T_{\tau_m\tau_m}T_{\tau_0\tau_0}^{1-m}
\end{align}
We use  \eqref{fusioneqs} to remove all of the fusion coefficients except one:
\begin{align}
\sum_{\l\in\Lambda_+^k}{\mbox{dim}}\l\, T_{\l\l} N_{\m\l}^\n={1\over T_{00}S_{00}}(TST)_{\m\n}
\end{align}
therefore:
\begin{align}
|X_L|&=\sum_{\s_0\tau_0}({\mbox{dim}}\s_0)^{1-n}({\mbox{dim}})^{1-m} N_{\tau\l}^{\s_0} T_{\s_0\s_0}^{-n-1} T_{\tau_0\tau_0}^{1-m} {1\over(T_{00}S_{00})^{n+m}} \times \nonumber \\
&\times (TST)_{\m_1\s_0}(TST)_{\m_2\s_0}\cdots(TST)_{\m_n\s_0}(TST)_{\n_1\tau_0}(TST)_{\n_2\tau_0} \cdots(TST)_{\n_m\tau_0}
\end{align}
Collecting the common factors of $T$ this equals:
\begin{align}
|X_L|&={T_{\m_1\m_1}\cdots T_{\m_n\m_n}T_{\n_1\n_1}\cdots T_{\n_m\n_m}\over(T_{00}S_{00})^{n+m}} \times\sum_{\s_0\tau_0}
({\mbox{dim}}\s_0)^{1-n}({\mbox{dim}}\tau_0)^{1-m} \times \nonumber \\
&\times N_{\tau_0\l}^{\s_0}T_{\s_0\s_0}^{-1}T_{\tau_0\tau_0}S_{\m_1\s_0}
\cdots S_{\m_n\s_0}S_{\n_1\tau_0}\cdots S_{\n_m\tau_0}
\end{align}
By filling in  the definition of the fusion coefficients and the ``quantum dimensions'' $\dim(\lambda)$
 we can rewrite this as
\begin{align}
|X_L|&={T_{\m_1\m_1}\cdots T_{\m_n\m_n}T_{\n_1\n_1}\cdots T_{\n_m\n_m}\over T_{00}^{n+m} S_{00}^2}
\sum_{\s_0\tau_0\s} (S_{0 \s_0})^{1-n} (S_{0 \tau_0})^{1-m} {S_{\s_0\s}^{-1}S_{\tau_0\s}S_{\l\s}\over S_{\s 0}}\times \nonumber \\
&\times T_{\s_0\s_0}^{-1}T_{\tau_0\tau_0} S_{\m_1\s_0}\cdots S_{\m_n\s_0}S_{\n_1\tau_0}\cdots S_{\n_m\tau_0}
\end{align}

\bigskip

For example, in the special case  $m=2,n=1$ we have
\begin{align}
|X_L|={T_{\m_1\m_1}T_{\n_1\n_1}T_{\n_2\n_2}\over T_{00}^3S_{00}^2}\sum_{\s_0\tau_0\s}{S_{\s_0\s}^{-1}S_{\tau_0\s}S_{\l\s}T_{\s_0\s_0}^{-1}T_{\tau_0\tau_0}S_{\m_1\s_0}S_{\n_1\tau_0}S_{\n_2\tau_0}\over S_{\tau_00}S_{\s 0}}.
\end{align}
which -- using $ST^{-1}S^{-1}=TST$ -- can be reduced to:
\begin{align}
|X_L|={T_{\m_1\m_1}^2T_{\n_1\n_1}T_{\n_2\n_2}\over T_{00}^3S_{00}^2}\sum_{\tau_0\s}{1\over S_{\tau_00}S_{\s 0}} S_{\tau_0\s}S_{\l\s}S_{\m_1\s}S_{\n_1\tau_0}S_{\nu_2\tau_0}T_{\s\s}T_{\tau_0\tau_0}.
\end{align}
\end{example}

\begin{example}  \label{Ex3} \rm
Note that $X_L$ is  also defined if $L$ is the ``empty'' link
$\emptyset$. In this case one has
 \be\label{genusg}
|X_{\emptyset}|=\sum_{\l \in \Lambda^k_+}
(\mbox{dim}\,\l)^{2-2g}~. \ee
 where $g$ is the genus of the  surface $\Sigma$.
\end{example}

 \section{State sums from the  Chern-Simons path integral in the torus gauge}
\label{sec4}

  \subsection{Chern-Simons models}
\label{subsec4.1}

  Let $M$ be an oriented compact 3-manifold
   and $\cA$ the space of smooth $\cG$-valued 1-forms on $M$.
   Without loss of generality  we can assume that the group $G$ fixed in Subsec.  \ref{subsec2.1} above is a Lie
subgroup of $U(N)$, $N \in \bN$. The Lie algebra $\cG$ of $G$ can
then be identified with the obvious Lie subalgebra of
 the Lie algebra $u(N)$ of $U(N)$.\par
     The   Chern-Simons  action function  $S_{CS}$ associated to $M$, $G$, $k$
     (with $k$  as in Subsec.   \ref{subsec2.2})
  is given by
$$S_{CS}(A) = \tfrac{k}{4\pi} \int_M \Tr(A \wedge dA + \tfrac{2}{3} A\wedge A\wedge A), \quad
A \in \cA$$
 with $\Tr:= c \cdot \Tr_{\Mat(N,\bC)}$ where
the normalization constant  $c$ is chosen\footnote{such a
normalization is always possible because by assumption $\cG$ is
simple so all $\Ad$-invariant scalar products on $\cG$ are
proportional to the Killing metric}
 such that\footnote{ Here ``$\cdot$'' is, of course, the standard
multiplication in $\Mat(N,\bC)$ and the wedge product $\wedge$
appearing in Eq. \eqref{eq_cond_on_G} is the one for
$\Mat(N,\bC)$-valued forms.}
\begin{equation} \label{eq_cond_on_G}
(A,B)  =   -\tfrac{1}{4 \pi^2}  \Tr(A \cdot B)    \quad \forall
A,B \in \cG
\end{equation}
holds, cf. e.g. \cite{Roz,Sawon} where the same normalization is used.

\begin{example}
If $G=SU(N)$ then $c=1$ so in this
case $\Tr$ coincides with $\Tr_{\Mat(N,\bC)}$.
\end{example}

From the definition of $S_{CS}$ it is obvious that $S_{CS}$ is invariant under
(orientation-preserving) diffeomorphisms.
Thus, at a heuristic level, we can expect that the
heuristic integral (the ``partition function'')
$Z(M) := \int  \exp(i S_{CS}(A)) DA$
is a topological invariant of the  3-manifold $M$.
Here $DA$ denotes  the informal ``Lebesgue measure'' on the space $\cA$.\par
Similarly,  we can expect that the mapping
which maps every sufficiently ``regular''
 colored link $L= ((l_1, l_2, \ldots, l_n),(\gamma_1,\gamma_2,\ldots,\gamma_n))$ in $M$
to the heuristic integral (the ``Wilson loop observable'' associated to $L$)
\begin{equation} \label{eq_WLO}
\WLO(L)  := \frac{1}{Z(M)} \int \prod_i  \Tr_{\rho_i}\bigl(\cP \exp\bigl(\int_{l_i} A\bigr)\bigr) \exp(i S_{CS}(A)) DA
\end{equation}
is a link invariant (or, rather, an invariant of colored links).
Here we have set $\rho_i:=\rho_{\gamma_i}$  $i \le n$,  (cf.
Subsec. \ref{subsec2.1}),
 $\Tr_{\rho_i}$ is the  trace in the representation $\rho_i$,
and $\cP \exp\bigl(\int_{l_i} A\bigr)$ denotes the
  holonomy of $A$ around the loop $l_i$.\par

Let us now consider the special case $M=\Sigma \times S^1$ where
$\Sigma$ is a closed oriented surface.
 Due to the well-known ``equivalence'' of Witten's invariants
and
 the Reshetikhin/Tureav invariants (cf., e.g., \cite{Walk})
 and the equivalence of the Reshetikhin/Tureav
 invariants with the shadow invariant (cf. Theorem 3.3 in Chap. X in \cite{turaevbook})
one can conclude that in this situation $\WLO(L)$ should coincide
with $|X_L|$ up to a multiplicative constant (independent of the
link). The value of this constant can be determined by looking at
the special case $L=\emptyset$, i.e. where $L$ is the ``empty'' link.
 As  $\WLO(\emptyset)= 1$ one can
conclude that $\WLO(L) =  \frac{1}{|X_{\emptyset}|} \cdot |X_L|$
should hold. One of the goals of this paper is to show this
formula directly (for the special situation where the link $L$
fulfills Assumptions \ref{assump1} and \ref{assump2} above) by
applying a suitable gauge fixing procedure to the Chern-Simons
path integral. This generalizes\footnote{in \cite{Ha4} only for
the case where $G=SU(2)$ and where each $\gamma_j$ was the highest
weight of the fundamental representation  the full path integral
was evaluated explicitly}
 the treatment in \cite{Ha4}.

\subsection{Torus gauge fixing applied to Chern-Simons models}
\label{subsec4.2}

In the present section and in Sec. \ref{subsec4.3} below we will
give a short summary of those results from \cite{Ha4} which will be relevant later.
Our presentation will not be totally self-contained, so the reader will probably find it helpful
to have a look at \cite{Ha4} for more details.

\medskip

 During the rest of this paper we will set
  $M := \Sigma \times S^1$ where  $\Sigma$ is a closed oriented surface.
Moreover, we will fix an arbitrary point $\sigma_0
\in \Sigma$ and an arbitrary\footnote{in order to simplify the
notation somewhat we will later restrict ourselves to  the special
case where  $t_0= i_{S^1}(0)=1$} point $t_0 \in S^1$.\par

 By  $\cA_{\Sigma}$ (resp.  $\cA_{\Sigma,\ct}$)
 we will denote the space of
smooth $\cG$-valued (resp.
  $\ct$-valued) 1-forms on $\Sigma$.
  $\tfrac{\partial}{\partial t}$ will denote
 the vector field on $S^1$ which is induced by the curve
 $i_{S^1}:[0,1] \ni t \mapsto e^{2\pi i t} \in S^1 \subset \bC$  and  $dt$  the  1-form
 on $S^1$ which is dual to   $\tfrac{\partial}{\partial t}$. We can lift $\tfrac{\partial}{\partial t}$ and $dt$
 in the obvious way to a vector field resp. a 1-form on $M$,
 which will also be denoted by $\tfrac{\partial}{\partial t}$ resp. $dt$.
  Every   $A \in \cA$ can be written uniquely in the form
  $A = A^{\orth} + A_0 dt $ with $A^{\orth} \in \cA^{\orth}$ and $A_0 \in C^{\infty}(M,\cG)$
  where $\cA^{\orth}$ is defined  by
 \begin{equation}
 \cA^{\orth} := \{A \in \cA \mid A(\tfrac{\partial}{\partial t}) =0 \}
 \end{equation}
We say that $A \in \cA$ is in the ``$T$-torus gauge'' if
$A \in  \cA^{\orth} \oplus  \{ B dt \mid B \in C^{\infty}(\Sigma,\ct)\}$.\par

By computing the relevant Faddeev-Popov determinant\footnote{cf. Sec. 2.3, Sec. 2.4,
 and Appendix C in the {\em latest}
version of \cite{Ha4}, i.e. [arXiv:math-ph/0507040v7].
We remark that the print version of  \cite{Ha4} contains an error in Sec. 2.3
(cf. footnote 10 in the latest version of \cite{Ha4}).
Moreover, Appendix C is missing in the print version of \cite{Ha4}}
  one obtains\footnote{
cf. Eq. (2.23) in Sec. 2.4 of \cite{Ha4} for a variant of this equation
where $C^{\infty}(\Sigma,P)$ instead of $C^{\infty}(\Sigma,\ct)$ appears.
Observe that Eq. (2.23) in \cite{Ha4} assumes $\Sigma$ to be non-compact.
The compact case is covered by Eq. (2.28) in \cite{Ha4}}
for every gauge-invariant function $\chi:\cA \to \bC$
\begin{equation}
 \int_{\cA} \chi(A) DA
 = const. \int_{C^{\infty}(\Sigma,\ct)} \biggl[ \int_{\cA^{\orth}} \chi(A^{\orth} + B dt)
   DA^{\orth}  \biggr] \det\bigl(1_{\ck}-\exp(\ad(B))_{| \ck}\bigr) DB
  \end{equation}
 where  $DA^{\orth}$ denotes the (informal) ``Lebesgue measure'' on $\cA^{\orth}$
and  $DB$ the (informal)  ``Lebesgue measure'' on  $C^{\infty}(\Sigma,\ct)$.\par

 In the special case  where $\chi(A) = \prod_i  \Tr_{\rho_i}\bigl(\cP \exp\bigl(\int_{l_i} A\bigr)\bigr)
 \exp(iS_{CS}(A))$
 we then get
 \begin{align} \label{eq_FADPOP_WLO} & \WLO(L)   \sim     \int_{C^{\infty}(\Sigma,\ct)}
 \biggl[ \int_{\cA^{\orth}}\prod_i  \Tr_{\rho_i}\bigl(\cP \exp\bigl(\int_{l_i} A^{\orth} + Bdt\bigr)\bigr)  \exp(iS_{CS}(A^{\orth} + Bdt)) DA^{\orth} \biggr] \nonumber \\
& \quad \times  \det\bigl(1_{\ck}-\exp(\ad(B))_{|\ck}\bigr) DB
\end{align}
Here and in the sequel $\sim$ denotes equality up to a
  multiplicative  constant independent of $L$.
Now
$$S_{CS}(A^{\orth} + B dt ) = \tfrac{k}{4\pi}  \int_M \bigl[ \Tr(A^{\orth} \wedge
dA^{\orth} ) + 2 \Tr( A^{\orth} \wedge  B dt \wedge
A^{\orth}) + 2  \Tr(A^{\orth} \wedge dB \wedge dt) \bigr]$$

so $S_{CS}(A^{\orth} + B dt )$ is quadratic in $A^{\orth}$ for
fixed $B$, which means that the informal (complex) measure $
\exp(i  S_{CS}( A^{\orth} + B dt)) DA^{\orth}$ appearing above is
of ``Gaussian type''. This increases the chances of
making rigorous sense of the right-hand side of Eq.
\eqref{eq_FADPOP_WLO} considerably.

\medskip

So far we have ignored the following two ``subtleties''
\begin{enumerate}
\item When one tries to find a rigorous meaning for the informal measure resp.
the corresponding integral functional in Eq. \eqref{eq_FADPOP_WLO} above
 one encounters certain problems   which can be solved
 by introducing a suitable decomposition\footnote{for a detailed motivation of this
decomposition, see Sec. 8  in \cite{Ha3b}}
$\cA^{\orth} = \hat{\cA}^{\orth} \oplus \cA^{\orth}_c$, which we
will describe now :\par

  Let us make the identification $\cA^{\orth} \cong
C^{\infty}(S^1,\cA_{\Sigma})$ where $C^{\infty}(S^1,\cA_{\Sigma})$
denotes the space of all ``smooth'' functions $\alpha:S^1 \to
\cA_{\Sigma}$, i.e. all functions $\alpha: S^1 \to \cA_{\Sigma}$
with the property that every smooth vector field  $X$  on $\Sigma$
the function $ \Sigma \times S^1 \ni (\sigma,t) \mapsto
\alpha(t)(X_{\sigma})$ is smooth.

The decomposition $\cA^{\orth} = \hat{\cA}^{\orth} \oplus \cA^{\orth}_c$ is defined by\footnote{note
 that the space $\hat{\cA}^{\orth}$ depends on the choice of the point $t_0$,
 and so will  some expressions  appearing later, see e.g.  Eq. \eqref{eq_1.int_carried_out} below}
\begin{align}
\hat{\cA}^{\orth} & := \{ A^{\orth} \in C^{\infty}(S^1,\cA_{\Sigma}) \mid \pi_{\cA_{\Sigma,\ct}}(A^{\orth}(t_0)) =0\},\\
 \cA_{c}^{\orth}  & :=   \{ A^{\orth} \in  C^{\infty}(S^1,\cA_{\Sigma}) \mid A^{\orth}   \text{ is constant and }\cA_{\Sigma,\ct}\text{-valued}\}
\end{align}
where $\pi_{\cA_{\Sigma,\ct}}: \cA_{\Sigma} \to \cA_{\Sigma,\ct}$
is the projection onto the first component in the decomposition
$\cA_{\Sigma} = \cA_{\Sigma,\ct} \oplus  \cA_{\Sigma,\ck}$.
It turns out that $S_{CS}$ behaves nicely under this decomposition.
More precisely, we have
\begin{equation}  \label{eq8.4new2}
  S_{CS}(\hat{A}^{\orth} + A^{\orth}_c + Bdt)  =  S_{CS}(\hat{A}^{\orth} + Bdt) + \frac{k}{2\pi} \int_{\Sigma} \Tr(dA^{\orth}_c \cdot B)
\end{equation}

Using this and  setting
$d\hat{\mu}^{\orth}_B(\hat{A}^{\orth})  :=\tfrac{1}{\hat{Z}(B)} \exp(i S_{CS}( \hat{A}^{\orth} + B dt))D\hat{A}^{\orth}
$ where  $\hat{Z}(B)  := \int \exp(i S_{CS}( \hat{A}^{\orth} + B dt))D\hat{A}^{\orth}$
we obtain
  \begin{multline} \label{eq_WLO_after_first_subtlety}
\WLO(L)  \sim
  \int_{C^{\infty}(\Sigma,\ct)}  \int_{\cA_c^{\orth}}  \biggl[ \int_{\hat{\cA}^{\orth}}
 \prod_i \Tr_{\rho_i}\bigl(\cP \exp\bigl(\int_{l_i}(\hat{A}^{\orth} + A^{\orth}_c  +  Bdt)  \bigr)\bigr)
   d\hat{\mu}^{\orth}_B(\hat{A}^{\orth}) \biggr] \\
\times  \exp( i \tfrac{k}{2\pi} \int_{\Sigma} \Tr(dA^{\orth}_c \cdot B)) DA_c^{\orth}
  \det\bigl(1_{\ck}-\exp(\ad(B)_{| \ck})\bigr)  \hat{Z}(B) \bigr\}  DB
 \end{multline}

A more careful analysis shows that  in the formula above one can
 replace $\ct$ by $\ct_{reg}$ or, alternatively, by
the Weyl alcove $P$. This amounts to including the extra factor
$1_{C^{\infty}(\Sigma,\ct_{reg})}$ or $1_{C^{\infty}(\Sigma,P)}$
in the integral expression above. In the sequel we will use the
factor $1_{C^{\infty}(\Sigma,P)}$.

\medskip

\item If one studies the torus gauge fixing procedure
  more closely for compact $\Sigma$ one finds that --
 due to certain topological obstructions (cf. \cite{BlTh3,Ha3c,Ha4}) --
 in general a 1-form $A$ can  be gauge-transformed into
a 1-form of the type $A^{\orth}+ Bdt$
only if one uses a gauge transformation $\Omega$
which has a certain  (mild) singularity
and if one allows $A^{\orth}$ to have a similar  singularity.
Concretely, in  \cite{Ha3c,Ha4} we worked
with  gauge transformations $\Omega$ of the type
$\Omega = \Omega_{smooth} \cdot \Omega_{\sing}(\cl) \in C^{\infty}((\Sigma \backslash \{\sigma_0\}) \times S^1,G)$
with $\Omega_{smooth} \in C^{\infty}(\Sigma  \times S^1,G)$
and $\Omega_{\sing}(\cl) \in C^{\infty}(\Sigma \backslash \{\sigma_0\},G) \subset
 C^{\infty}((\Sigma \backslash \{\sigma_0\}) \times S^1,G)$
where $\sigma_0 \in \Sigma$ is the point fixed above and where the
parameter
 $\cl$  is an element of $ [\Sigma,G/T]$, i.e.
a homotopy class of mappings from $\Sigma$ to $G/T$.
$\Omega_{\sing}(\cl)$ is obtained from $\cl$  by fixing a
representative $\bar{g}(\cl) \in C^{\infty}(\Sigma,G/T)$ of $\cl$
and then lifting the restriction
 $\bar{g}(\cl)_{|\Sigma \backslash \{\sigma_0\}}: \Sigma \backslash \{\sigma_0\} \to G/T$
 to a mapping $\Sigma \backslash \{\sigma_0\} \to G$.
 In other words:
$\Omega_{\sing}(\cl) \in C^{\infty}(\Sigma \backslash \{\sigma_0\},G)$
 is a fixed mapping with the property  that
 $\pi_{G/T} \circ \Omega_{\sing}(\cl) = \bar{g}(\cl)_{|\Sigma \backslash \{\sigma_0\}}$
 where $\pi_{G/T} : G \to G/T$ is the canonical projection.\par
 The use of the singular gauge transformations $\Omega_{\sing}(\cl)$
 gives rise to an extra summation  $\sum_{\cl \in [\Sigma,G/T]}$
 and to extra terms $A^{\orth}_{sing}(\cl):= \pi_{\ct}(\Omega_{\sing}(\cl)^{-1} \cdot d\Omega_{\sing}(\cl))$,
 i.e. in Eq. \eqref{eq_WLO_after_first_subtlety} above we have to include a summation  $\sum_{\cl \in [\Sigma,G/T]}$
 and we have to replace  $A^{\orth}_c$ by $A^{\orth}_c+ A^{\orth}_{sing}(\cl)$
 (for a detailed description and justification of all this, see
 \cite{Ha3c,Ha4}).
\end{enumerate}

\noindent Taking into account these two subtleties we obtain
  \begin{multline}  \label{eq_WLO_0}
\WLO(L)  \sim\\
 \sum_{\cl \in [\Sigma,G/T]} \int_{\cA_c^{\orth} \times C^{\infty}(\Sigma,\ct)}  1_{C^{\infty}(\Sigma,P)}(B)  \biggl[ \int_{\hat{\cA}^{\orth}}
  \prod_i \Tr_{\rho_i}\bigl(\cP \exp\bigl(\int_{l_i}(\hat{A}^{\orth} + A^{\orth}_c  + A^{\orth}_{sing}(\cl) +  Bdt)
  \bigr)\bigr)    d\hat{\mu}^{\orth}_B(\hat{A}^{\orth}) \biggr] \\
\times \bigl\{ \exp( i \tfrac{k}{2\pi} \int_{\Sigma}  \Tr(dA^{\orth}_{sing}(\cl) \cdot B)  )
 \det\bigl(1_{\ck}-\exp(\ad(B)_{| \ck})\bigr)  \hat{Z}(B) \bigr\}\\
  \times \exp( i \tfrac{k}{2\pi} \int_{\Sigma} \Tr(dA^{\orth}_c \cdot B)) (DA_c^{\orth} \otimes    DB)
 \end{multline}
 where
$$ \int_{\Sigma}  \Tr(dA^{\orth}_{sing}(\cl) \cdot B)
   :=  \lim_{\eps \to 0} \int_{\Sigma \backslash B_{\eps}(\sigma_0)}
 \Tr(dA^{\orth}_{sing}(\cl) \cdot B)
$$
Here $B_{\eps}(\sigma_0)$ is the closed $\eps$-ball around
$\sigma_0$ with respect to  an arbitrary but fixed Riemannian
metric on $\Sigma$.

 \begin{remark}   \label{rem4} \rm
The mapping $n: [\Sigma,G/T] \to \ct$  given  by $n(\cl) =
\lim_{\eps \to 0} \int_{\Sigma \backslash B_{\eps}(\sigma_0)}
d A^{\orth}_{sing}(\cl)$ is independent of the special choice of
 $\bar{g}(\cl)$ and $\Omega_{sing}(\cl)$, cf.  \cite{Ha3c,Ha4}.
Moreover, and this will be important in Subsec. \ref{subsec4.4}
below one can show that $n$ is  a bijection from $[\Sigma,G/T]$
onto $I= \ker(\exp_{| \ct})$ (cf. also \cite{BlTh3} for a similar result).
 \end{remark}

  \subsection{Some comments regarding a rigorous realization of the r.h.s. of Eq. \eqref{eq_WLO_0}}
\label{subsec4.3}

In \cite{Ha4,Ha6} it is explained how one can make
rigorous sense of the  path integral expression appearing on the right-hand side
of Eq. \eqref{eq_WLO_0}
using results/constructions from White Noise Analysis (cf., e.g., \cite{HKPS}),
   certain regularization techniques like ``loop smearing'' and ``framing'', and a
   suitable  regularization of the expression $\det\bigl(1_{\ck}-\exp(\ad(B)_{| \ck})\bigr)  \hat{Z}(B)$ appearing above.  We do not want to repeat the discussion
  in \cite{Ha4,Ha6} in the present paper.
  Let us just remark the following:

\begin{enumerate}
\item In view of the results \cite{BlTh1} it is clear how to
make sense of the factor $\det\bigl(1_{\ck}-\exp(\ad(B)_{|
\ck})\bigr)   \hat{Z}(B)$  appearing  Eq. \eqref{eq_WLO_0} in the special case where $B$ is a constant
function (this was the only case which was relevant in
\cite{BlTh1}). More precisely, using the same
 $\zeta$-function regularization as the one described  in Sec. 6 in \cite{BlTh1}
 one comes to the conclusion that in this special case of constant $B \equiv b$ the expression
$\det\bigl(1_{\ck}-\exp(\ad(B)_{| \ck})\bigr) \hat{Z}(B)$ should be replaced by\footnote{
the first factor in Eq. \eqref{expr_reg_det}
 gives rise to the so-called ``charge shift'' $k \to k + c_G$.
Let us mention that the claim that such a charge shift appears is
contested by some authors, cf. Remark B.2 in \cite{Ha7b}.
If one does not believe that such a  charge shift will appear
one will have to omit the first factor  in Eq. \eqref{expr_reg_det}
and the analogous terms in the equations below}

\begin{equation} \label{expr_reg_det}
\exp( i \tfrac{\cg}{2\pi} \int_{\Sigma} \Tr(dA^{\orth}_c \cdot B))
\times \det\nolimits_{reg}\bigl(1_{\ck}-\exp(\ad(B)_{| \ck})\bigr)
\end{equation}
where
$$\det\nolimits_{reg}\bigl(1_{\ck}-\exp(\ad(B)_{| \ck})\bigr):=
\det\bigl(1_{\ck}-\exp(\ad(b))_{|\ck}\bigr)^{\chi(\Sigma)/2}$$
In \cite{Ha4} it was suggested that in the more general situation
 where $B$ is a  step function of the form  $ B=\sum_{t=0}^{\mu} b_t 1_{Y_t}$
 one should again replace $\det\bigl(1_{\ck}-\exp(\ad(B)_{| \ck})\bigr) \hat{Z}(B)$  by expression \eqref{expr_reg_det} where now
 \begin{equation} \label{eq_detreg}
\det\nolimits_{reg}\bigl(1_{\ck}-\exp(\ad(B)_{| \ck})\bigr) :=
\prod_{t=0}^{\mu}
\det\bigl(1_{\ck}-\exp(\ad(b_t))_{|\ck}\bigr)^{\chi(Y_t)/2}
\end{equation}
Moreover, it was suggested that one should include a $\exp( i
\tfrac{\cg}{2\pi} \int_{\Sigma} \Tr(dA^{\orth}_{sing}(\cl) \cdot
B))$-factor in the expression  \eqref{expr_reg_det} above.
 Incorporating these changes  into \eqref{eq_WLO_0} one obtains
\begin{multline} \label{eq_WLO_1}
\WLO(L)  \sim\\
 \sum_{\cl \in [\Sigma,G/T]} \int_{\cA_c^{\orth} \times C^{\infty}(\Sigma,\ct)}  1_{C^{\infty}(\Sigma,P)}(B)
 \biggl[  \int_{\hat{\cA}^{\orth}} \prod_i \Tr_{\rho_i}\bigl(\cP \exp\bigl(\int_{l_i}(\hat{A}^{\orth} + A^{\orth}_c
 + A^{\orth}_{sing}(\cl) +  Bdt)  \bigr)\bigr)    d\hat{\mu}^{\orth}_B(\hat{A}^{\orth}) \biggr] \\
\times \bigl\{ \exp( i \tfrac{k+\cg}{2\pi} \int_{\Sigma}
\Tr(dA^{\orth}_{sing}(\cl) \cdot B)  )
\det\nolimits_{reg}\bigl(1_{\ck}-\exp(\ad(B)_{| \ck})\bigr)
\bigr\}   d\nu((A_c^{\orth},B))
 \end{multline}
where we have introduced  the heuristic complex measure $d\nu$
given by    $$d\nu((A_c^{\orth},B)):=
   \exp( i \tfrac{k+\cg}{2\pi} \int_{\Sigma} \Tr(dA^{\orth}_c \cdot B)) (DA_c^{\orth} \otimes
   DB)$$

\item The ``Gauss type'' integral functionals  $\int_{\hat{\cA}^{\orth}}\cdots
d\hat{\mu}^{\orth}_B(\hat{A}^{\orth})  $ resp.
   $\int_{\cA_c^{\orth} \times C^{\infty}(\Sigma,\ct)} \cdots d\nu((A_c^{\orth},B))$
  appearing above     can be realized rigorously as  Hida
  distributions $\Phi^{\orth}_B$ resp. $\Psi$
     on suitable extensions of the spaces $\hat{\cA}^{\orth}$ and $\cA_c^{\orth} \times
     C^{\infty}(\Sigma,\ct)$.
Moreover, also the space $C^{\infty}(\Sigma,P)$ appearing in the
indicator function $ 1_{C^{\infty}(\Sigma,P)}$ must be replaced by
a larger space. (The fact that one has to extend the
      original spaces of smooth functions by larger spaces
     consisting of less regular functions is a  usual phenomenon
     in Constructive Quantum Field Theory.)
The details regarding the  extensions of the spaces
$\hat{\cA}^{\orth}$ and $\cA_c^{\orth} \times
     C^{\infty}(\Sigma,\ct)$ have been or will be  discussed elsewhere\footnote{
     the extension of $\hat{\cA}^{\orth}$
     is described in Sec. 8 in \cite{Ha3b}, see also Sec. 4 in
     \cite{Ha4};      the extension of $\cA_c^{\orth} \times
     C^{\infty}(\Sigma,\ct)$  was described in \cite{Ha6} }
     and they are not relevant if one is only interested in a heuristic
    evaluation of the r.h.s of Eq. \eqref{eq_WLO_0} resp.
    \eqref{eq_WLO_1}. By contrast, the question of how to extend the space
$C^{\infty}(\Sigma,P)$  appearing in the indicator function $
1_{C^{\infty}(\Sigma,P)}$ is more subtle even if one is only
interested in a heuristic treatment. One might think that
 if one replaces $C^{\infty}(\Sigma,P)$ by the space $P^{\Sigma}$
 of {\em all} $P$-valued
  functions on $\Sigma$ this should be enough.
  In fact that was the ansatz used in \cite{Ha4}
and in the special case where all the link colors $\gamma_i$ are
(minimal) fundamental weights this ansatz works. However, it turns
out that in the case of general link colors $\gamma_i$ the space
$P^{\Sigma}$ is too small. In order to find the ``correct'' space
note that $1_{C^{\infty}(\Sigma,P)}(B)=
1_{C^{\infty}(\Sigma,\ct_{reg})}(B) 1_{P}(B(\sigma_0))$. This
 suggests that one might try to replace
  $C^{\infty}(\Sigma,\ct_{reg})$ by
$(\ct_{reg})^{\Sigma}$.
  As the computations in the next subsections show
 the second ansatz is the ''correct'' one.
Of course, it would be desirable to find a thorough justification
for using the second ansatz which is independent of the results in
the rest of this paper.

\item For the implementation of the ``framing procedure''  in
\cite{Ha4,Ha6} a suitable family $(\phi_s)_{s>0}$ of diffeomorphisms
of $\Sigma \times S^1$ fulfilling certain condition (see list
below) was fixed. For  each diffeomorphism $\phi_s$   a
``deformation'' $\Phi^{\orth}_{B,\phi_s}$ resp.
$\Psi_{\bar{\phi}_s}$ of $\Phi^{\orth}_B$ resp. $\Psi$ was then
introduced and used to replace $\Phi^{\orth}_B$ and $\Psi$
 in the original formula. Later the free parameter $s >0$
in the resulting formulas was eliminated by taking the limit $s
\to 0$.\par

Among others $(\phi_s)_{s>0}$ was
 assumed to fulfill  the following conditions
\begin{itemize}

\item $\phi_s \to \id_M  \text{ as   } s \to 0$ uniformly w.r.t. to
an arbitrary Riemannian metric on $M$.

\item $(\phi_s)^* \cA^{\orth} = \cA^{\orth}$ for all $s>0$. This
 condition implies that each $\phi_s$, $s
>0$, is of the form
$$\phi_s(\sigma,t)= (\bar{\phi}_s(\sigma), v_s(\sigma,t)) \quad \forall (\sigma,t) \in \Sigma \times S^1$$
for a uniquely determined diffeomorphism $\bar{\phi}_s:\Sigma \to
\Sigma$ and $v_s \in C^{\infty}(\Sigma \times S^1,S^1)$.

\item  $(\phi_s)_{s>0}$ is ``horizontal'' in the sense
that\footnote{in fact the definition of the term ``horizontal'' in
 \cite{Ha4} was somewhat broader but also more complicated} it can be obtained by integrating
a smooth vector field $X$ on $\Sigma \times S^1$, which for all $i
\le n$, $u \in [0,1]$ is  orthogonal to the tangent vector
$l'_i(u)$ (i.e. $X(l_i(u)) \perp l'_i(u)$) and, at the same time,
horizontal in  $l_i(u)$ (i.e. $dt(X(l_i(u)))=0$).

\end{itemize}

\end{enumerate}

\subsection{Explicit heuristic evaluation of the WLOs}
\label{subsec4.4}

As mentioned above we will not go into details concerning a
rigorous realization of the r.h.s. of \eqref{eq_WLO_1} but give a
short heuristic treatment instead. As the starting point for this
heuristic treatment we use the following
modification\footnote{clearly, the modification consists in
replacing the integration $\int \cdots d\nu$
 by the two separate  integrations $\int \cdots DB$ and
 $\int_{\cA_c^{\orth}} \cdots DA_c^{\orth}$
and in the use of $1_{(\ct_{reg})^{\Sigma}}(B) \
1_{P}(B(\sigma_0))$ instead of $1_{C^{\infty}(\Sigma,P)}(B)=
1_{C^{\infty}(\Sigma,\ct_{reg})}(B) 1_{P}(B(\sigma_0))$} of Eq.
\eqref{eq_WLO_1} above.
\begin{multline} \label{eq_WLO_2}
\WLO(L)  \sim\\
 \sum_{\cl \in [\Sigma,G/T]} \int 1_{(\ct_{reg})^{\Sigma}}(B) 1_P(B(\sigma_0))
 \biggl[  \int_{\cA_c^{\orth}} \int_{\hat{\cA}^{\orth}} \prod_i \Tr_{\rho_i}\bigl(\cP \exp\bigl(\int_{l_i}(\hat{A}^{\orth} + A^{\orth}_c
 + A^{\orth}_{sing}(\cl) +  Bdt)  \bigr)\bigr)    d\hat{\mu}^{\orth}_B(\hat{A}^{\orth})  \\
\times \bigl\{ \exp( i \tfrac{k+\cg}{2\pi} \int_{\Sigma}  \Tr(dA^{\orth}_{sing}(\cl) \cdot B)  ) \det\nolimits_{reg}\bigl(1_{\ck}-\exp(\ad(B)_{| \ck})\bigr) \bigr\}\\
  \times \exp( i \tfrac{k+\cg}{2\pi} \int_{\Sigma} \Tr(dA^{\orth}_c \cdot B)) DA_c^{\orth} \biggr]   DB
 \end{multline}
   Let, for fixed $j \le n$,     $u_1, u_2, \ldots u_{n^j}$ be the ``solutions'' of the equation  $l^j_{S^1}(u)=t_0$,
    i.e.    those curve parameters
in which $l^j_{S^1}$ ``hits'' $t_0$.
For $m\le n^j$ we set
$\sigma^j_m  := l^j_{\Sigma}(u_m)$, and $\eps^j_m  := 1$ resp.
$\eps^j_m  := -1$ resp. $\eps^j_m  := 0$
  if \ $l^j_{S^1}$ \ crosses $t_0$ in $u_m$ ``from below''
  resp. ``from above'' resp.  only touches  $t_0$ in $u_m$.\par

In \cite{Ha4} it is shown how one can  evaluate
 (the rigorous realization of) the heuristic expression
$$\int_{\hat{\cA}^{\orth}} \prod_i \Tr_{\rho_i}\bigl(\cP \exp\bigl(\int_{l_i}(\hat{A}^{\orth} + A^{\orth}_c
 + A^{\orth}_{sing}(\cl) +  Bdt)  \bigr)\bigr)    d\hat{\mu}^{\orth}_B(\hat{A}^{\orth})$$
 explicitly (for links fulfilling  Assumption 1 and 2)
   and that by doing so one obtains the expression
$$ \prod_{j=1}^n \Tr_{\rho_j} \bigg[ \exp(\int_{l^j_{\Sigma}}
A^{\orth}_c)    \exp(\int_{l^j_{\Sigma}}   A^{\orth}_{sing}(\cl))
\exp(\ \sum\nolimits_{m=1}^{n^j} \eps^j_m  B(\sigma^j_m))  \biggr] $$

By plugging the last expression into Eq. \eqref{eq_WLO_2} we obtain
\begin{multline} \label{eq_1.int_carried_out}
\WLO(L)  \sim\\
 \sum_{\cl} \int 1_{(\ct_{reg})^{\Sigma}}(B) 1_P(B(\sigma_0)) \biggl[  \int_{\cA_c^{\orth}}  \prod_{j=1}^n \Tr_{\rho_j} \bigg[ \exp(\int_{l^j_{\Sigma}}   A^{\orth}_c)    \exp(\int_{l^j_{\Sigma}}   A^{\orth}_{sing}(\cl))
  \exp(\ \sum\nolimits_{m=1}^{n^j} \eps^j_m  B(\sigma^j_m)) \biggr] \\
\times \bigl\{ \exp( i \tfrac{k+\cg}{2\pi} \int_{\Sigma}  \Tr(dA^{\orth}_{sing}(\cl) \cdot B)  ) \det\nolimits_{reg}\bigl(1_{\ck}-\exp(\ad(B)_{| \ck})\bigr) \bigr\}\\
  \times \exp( i \tfrac{k+\cg}{2\pi} \int_{\Sigma} \Tr(dA^{\orth}_c \cdot B)) DA_c^{\orth} \biggr]   DB
 \end{multline}

Let us now fix an auxiliary Riemannian metric $\mathbf g$ on
$\Sigma$ for the rest of this paper. Let $\mu_{\mathbf g}$ denote
the corresponding volume measure  on $\Sigma$ and $\star$ the
Hodge star operator induced by $\mathbf g$. Moreover, let
$L^2_{\ct}(\Sigma,d\mu_{\mathbf g})$  denote obvious\footnote{ the
inner product $\ll \cdot, \cdot
\gg_{L^2_{\ct}(\Sigma,d\mu_{\mathbf g})}$ is given by $\ll B_1,
 B_2 \gg_{L^2_{\ct}(\Sigma,d\mu_{\mathbf g})} = \int_{\Sigma} (B_1(\sigma),B_2(\sigma)) d\mu_{\mathbf
 g}(\sigma)$
 where $(\cdot,\cdot)$ is the inner product on $\cG \supset \ct$
 fixed above.
 }
$L^2$-space. Then we have (cf. Eq. \eqref{eq_cond_on_G})
 $$ \int_{\Sigma} \Tr(d A^{\orth}_c \cdot B) =
 \int \Tr(\star d A^{\orth}_c \cdot B) d\mu_{\mathbf g} = - 4 \pi^2 \ll \star d A^{\orth}_c,
  B\gg_{L^2_{\ct}(\Sigma,d\mu_{\mathbf g})}$$
 From Stokes' Theorem we obtain
$$\int_{l^j_{\Sigma}}   A^{\orth}_c = \int_{R^+_j} d A^{\orth}_c = \int_{R^+_j} \star d A^{\orth}_c d\mu_{\mathbf g} =
\int \star d A^{\orth}_c \cdot  1_{R^+_j} d\mu_{\mathbf g}
$$
which implies
\begin{equation} \label{eq_1_R+}
 ( \alpha,  \int_{l^j_{\Sigma}}   A^{\orth}_c ) = \ll \star d A^{\orth}_c,
   \alpha \cdot   1_{R^+_j}      \gg_{L^2_{\ct}(\Sigma,d\mu_{\mathbf g})}
 \end{equation}
for every $\alpha \in \ct$. Here $1_{R^+_j}$ denotes the indicator
function of the region $R^+_j$ defined in Subsec. \ref{subsec3.1}
above. Note that Eq. \eqref{eq_1_R+} also holds if we replace
$1_{R^+_j}$ by
\begin{equation} \label{eq_1_shift}
1^{\shift}_{R^+_j}:= 1_{R^+_j} - 1_{R^+_j}(\sigma_0)
\end{equation}
We will use this modified version of Eq. \eqref{eq_1_R+} in the
sequel.
Finally, we take into account that
\begin{equation} \label{eq_multiplicities_mod}
\Tr_{\rho_j}(\exp(b))= \chi_{\rho_j}(\exp(b)) =
 \sum_{\alpha \in \Lambda} m_{\gamma_j}(\alpha) e^{2 \pi i (\alpha,b)} \quad \quad \forall b \in \ct
 \end{equation}
for suitable\footnote{$m_{\gamma_j}(\alpha)$ is just the multiplicity
of the weight $\alpha$ in the character $\chi_{\rho_j}$}
  $m_{\gamma_j}(\alpha) \in \bN_0$,   cf. Eq. \eqref{eq_multiplicities} above.    Setting
  \begin{equation}
\bar{\alpha} := 2 \pi \alpha,
\end{equation}
for each $\alpha \in \Lambda$ we  obtain from Eq. \eqref{eq_multiplicities_mod}
 and the modified version of Eq.  \eqref{eq_1_R+}
 \begin{multline*} \Tr_{\rho_j}\bigg[ \exp(\int_{l^j_{\Sigma}}   A^{\orth}_c)   \exp(\int_{l^j_{\Sigma}}   A^{\orth}_{sing}(\cl))
    \exp(\ \sum\nolimits_{m} \eps^j_m  B(\sigma^j_m))  \biggr]\\
     =   \sum_{\alpha \in \Lambda} m_{\gamma_j}(\alpha)   \exp( i
     ( \bar{\alpha}, \sum\nolimits_{m} \eps^j_m  B(\sigma^j_m)) )) \cdot
       \exp(i \int_{l^j_{\Sigma}}   (\bar{\alpha}, A^{\orth}_{sing}(\cl) ))
          \cdot  \exp( i  \ll \star d A^{\orth}_c,    \bar{\alpha} \cdot   1^{\shift}_{R^+_j}      \gg_{L^2_{\ct}(\Sigma,d\mu_{\mathbf g})})
\end{multline*}
(Here $(\bar{\alpha}, A^{\orth}_{sing}(\cl))$ denotes the obvious real-valued 1-form).
Plugging this into Eq. \eqref{eq_1.int_carried_out} above we get
\begin{align} \label{eq_reduce_WLO_sum}
& \WLO(L) \nonumber \\
& \sim  \sum_{\cl}   \int 1_{(\ct_{reg})^{\Sigma}}(B)
1_P(B(\sigma_0))  \int\limits_{\cA_c^{\orth}}
   \prod_{j=1}^n\bigg[\sum_{\alpha_j \in \Lambda} m_{\gamma_j}(\alpha_j) \exp(i \! \! \int_{l^j_{\Sigma}}
     ( \bar{\alpha}_j, A^{\orth}_{sing}(\cl) ))       \exp(i \! \!  \ll \! \star d A^{\orth}_c,    \bar{\alpha}_j    1^{\shift}_{R^+_j}
        \gg_{L^2_{\ct}(\Sigma,d\mu_{\mathbf g})}) \nonumber \\
   & \quad \quad \quad \times \exp( i  ( \bar{\alpha}_j, \sum\nolimits_{m} \eps^j_m  B(\sigma^j_m)))
    \biggr] \det\nolimits_{reg}\bigl(1_{\ck}-\exp(\ad(B)_{| \ck})\bigr)
    \exp( i \tfrac{k+\cg}{2\pi} \int_{\Sigma}  \Tr(dA^{\orth}_{sing}(\cl) \cdot B)  )  \nonumber \\
&   \quad \quad \quad \times
 \exp( - 2 \pi i (k+\cg)  \ll \star dA^{\orth}_c, B \gg_{L^2_{\ct}(\Sigma,d\mu_{\mathbf g})}) DA_c^{\orth} DB\nonumber \\
& = \sum_{\cl}  \sum_{\alpha_1,  \ldots, \alpha_n \in
\Lambda}   \bigl(\prod_{j=1}^n m_{\gamma_j}(\alpha_j)\bigr)
 \int 1_{(\ct_{reg})^{\Sigma}}(B) 1_P(B(\sigma_0))   \prod_{j=1}^n   \exp(i  \! \! \int_{l^j_{\Sigma}}   ( \bar{\alpha}_j, A^{\orth}_{sing}(\cl) ))   \exp( i
     ( \bar{\alpha}_j, \sum\nolimits_{m} \eps^j_m  B(\sigma^j_m)))  \nonumber \\
 & \quad \quad \quad \times  \det\nolimits_{reg}\bigl(1_{\ck}-\exp(\ad(B)_{| \ck})\bigr)
   \exp( i \tfrac{k+\cg}{2\pi} \int_{\Sigma}  \Tr(dA^{\orth}_{sing}(\cl) \cdot B)  )  \nonumber \\
& \quad \quad \quad \times \biggl[\int\limits_{\cA_c^{\orth}}
    \exp( i  \ll \star d A^{\orth}_c,   \sum_{j=1}^n  \bar{\alpha}_j \cdot   1^{\shift}_{R^+_j}
   -    2 \pi (k+\cg)  B \gg_{L^2_{\ct}(\Sigma,d\mu_{\mathbf g})}) DA_c^{\orth}\biggr] DB \nonumber \\
   & \overset{(*)}{=} \sum_{\cl}  \sum_{\alpha_1, \alpha_2, \ldots, \alpha_n \in \Lambda}   \bigl(\prod_{j=1}^n m_{\gamma_j}(\alpha_j)\bigr)
 \int 1_{(\ct_{reg})^{\Sigma}}(B) 1_P(B(\sigma_0))
  \exp( i \tfrac{k+\cg}{2\pi} \int_{\Sigma}  \Tr(dA^{\orth}_{sing}(\cl) \cdot B)  ) \nonumber \\
 & \quad \quad \quad \times  \det\nolimits_{reg}\bigl(1_{\ck}-\exp(\ad(B)_{| \ck})\bigr) \prod_{j=1}^n  \biggl(    \exp(i \int_{l^j_{\Sigma}}   (  \bar{\alpha}_j, A^{\orth}_{sing}(\cl) ))  \exp( i
     (  \bar{\alpha}_j, \sum\nolimits_{m} \eps^j_m  B(\sigma^j_m)) ))\biggr) \nonumber \\
& \quad \quad \quad  \times \biggl[ \delta\bigl( d\bigl(
\sum_{j=1}^n  \bar{\alpha}_j \cdot  1^{\shift}_{R^+_j}
   -  2 \pi (k+\cg) B \bigr) \bigr) \biggr] DB
\end{align}
   Here, in step $(*)$ we have used the informal equation
   \begin{multline} \label{eq_delta_repl}
 \int\limits_{\cA_c^{\orth}}
    \exp( i  \ll \star d A^{\orth}_c,   \sum_{j=1}^n  \bar{\alpha}_j \cdot   1^{\shift}_{R^+_j}
   -  2 \pi (k+\cg) B  \gg_{L^2_{\ct}(\Sigma,d\mu_{\mathbf g})})
   DA_c^{\orth}\\
  \sim \delta\bigl( d\bigl(  \sum_{j=1}^n  \bar{\alpha}_j \cdot   1^{\shift}_{R^+_j}
   -  2 \pi (k+\cg) B \bigr) \bigr)
\end{multline}
which is a kind of infinite dimensional analogue
 of the well-known informal
equation $ \int_{\bR} \exp(i\langle x,y \rangle) dx \sim
\delta(y)$. In fact, as $\bigl( \sum_{j=1}^n \bar{\alpha}_j \cdot
1^{\shift}_{R^+_j}   -  2 \pi (k+\cg) B \bigr)$ is in general not
smooth (not even continuous) we should be a little bit more
careful. Instead of using the delta-function $\delta\bigl( d\bigl(
\sum_{j=1}^n  \bar{\alpha}_j \cdot 1^{\shift}_{R^+_j}
   -  2 \pi (k+\cg) B \bigr) \bigr)$ in Eqs. \eqref{eq_reduce_WLO_sum} and \eqref{eq_delta_repl} above
  we should rather use  the  ``superposition''\footnote{one can
  equally well use the superposition
$\int_{\ct}  \cdots \delta\bigl(B-
   \tfrac{1}{ 2 \pi (k+\cg)}  \bigl(-b+ \sum_{j=1}^n  \bar{\alpha}_j \cdot 1^{\shift}_{R^+_j})  \bigr) \bigr) db$
   or $\int_{\ct}  \cdots \delta\bigl(B-
   \tfrac{1}{ 2 \pi (k+\cg)}  \bigl(b + \sum_{j=1}^n  \bar{\alpha}_j \cdot 1^{\shift}_{R^+_j})  \bigr) \bigr)  db$
the final result will be the same,
which is not surprising since, heuristically,
$\delta\bigl( d\bigl(  \sum_{j=1}^n  \bar{\alpha}_j \cdot   1^{\shift}_{R^+_j}    -  2 \pi (k+\cg) B \bigr) \bigr)
    \sim \delta\bigl( d\bigl(
     2 \pi (k+\cg) B  -  \sum_{j=1}^n  \bar{\alpha}_j \cdot   1^{\shift}_{R^+_j} \bigr) \bigr)
     \sim \delta\bigl( d\bigl(
     B  -  \tfrac{1}{ 2 \pi (k+\cg)} \sum_{j=1}^n  \bar{\alpha}_j \cdot   1^{\shift}_{R^+_j} \bigr) \bigr)$
}
   $$\int_{\ct}  \cdots \delta\bigl(B-
   \bigl(b+ \tfrac{1}{ 2 \pi (k+\cg)} \sum_{j=1}^n  \bar{\alpha}_j \cdot 1^{\shift}_{R^+_j}\bigr)  \bigr) \bigr)  db$$
   of delta-functions. Then we  obtain
\begin{align*}
& \WLO(L) \\
& \sim
  \sum_{\cl \in [\Sigma,G/T]}  \sum_{\alpha_1, \alpha_2, \ldots, \alpha_n \in \Lambda}   \bigl(\prod_{j=1}^n m_{\gamma_j}(\alpha_j)\bigr)
   \int_{\ct} \ db  \biggl[ \exp( i \tfrac{k+\cg}{2\pi} \int_{\Sigma}  \Tr(dA^{\orth}_{sing}(\cl) \cdot b)  )\\
 & \quad   \times    \biggl( 1_{(\ct_{reg})^{\Sigma}}(B) 1_P(B(\sigma_0)) \det\nolimits_{reg}\bigl(1_{\ck}-\exp(\ad(B)_{| \ck})\bigr)   \prod_{j=1}^n    \exp( i
     ( \bar{\alpha}_j, \sum\limits_{m} \eps^j_m  B(\sigma^j_m)) ) \biggr)_{| B=b +  \tfrac{1}{2 \pi (k+\cg)} \sum\limits_{j=1}^n  \bar{\alpha}_j    1^{\shift}_{R^+_j}}  \\
    & \quad  \times  \biggl\{  \exp(i \sum_j \int_{l^j_{\Sigma}}   (  \bar{\alpha}_j, A^{\orth}_{sing}(\cl) ))
     \exp( i \tfrac{k+\cg}{2\pi}   \int_{\Sigma} \Tr\bigl(dA^{\orth}_{sing}(\cl) \cdot
     \tfrac{1}{2 \pi (k+\cg)} \sum_{j=1}^n  \bar{\alpha}_j   1^{\shift}_{R^+_j}\bigr)\biggr\} \biggr] \\
&  \overset{(**)}{=}
  \sum_{\alpha_1,  \ldots, \alpha_n \in \Lambda}   \bigl(\prod_{j=1}^n m_{\gamma_j}(\alpha_j)\bigr)
  \sum_{\cl \in [\Sigma,G/T]}   \int_{\ct} \ db  \biggl[  \exp( i 2 \pi (k+\cg) ( n(\cl),  b ))\\
   & \quad  \times  \biggl( 1_{(\ct_{reg})^{\Sigma}}(B) 1_P(B(\sigma_0))   \det\nolimits_{reg}\bigl(1_{\ck}-\exp(\ad(B)_{| \ck})\bigr)  \prod_{j=1}^n    \exp( i
     ( \bar{\alpha}_j, \sum\limits_{m} \eps^j_m  B(\sigma^j_m)) ) \biggr)_{| B=b +  \tfrac{1}{2 \pi (k+\cg)} \sum\limits_{j=1}^n  \bar{\alpha}_j    1^{\shift}_{R^+_j}}  \\
& \quad \ \times \biggl\{ 1 \biggr\} \biggr]\\
&  \overset{(***)}{=}   \sum_{\alpha_1, \alpha_2, \ldots, \alpha_n \in \Lambda}   \bigl(\prod_{j=1}^n m_{\gamma_j}(\alpha_j)\bigr)
   \sum\nolimits_{b \in \frac{1}{ k+\cg} \Lambda}
 \biggl(1_{(\ct_{reg})^{\Sigma}}(B) 1_P(B(\sigma_0))  \\
 & \quad \quad \quad \times \det\nolimits_{reg}\bigl(1_{\ck}-\exp(\ad(B)_{| \ck})\bigr)  \prod_{j=1}^n
 \exp(2\pi i
     ( \alpha_j, \sum\nolimits_{m} \eps^j_m  B(\sigma^j_m)) ) \biggr)_{| B=b +  \tfrac{1}{ k+\cg} \sum_{j=1}^n  \alpha_j \cdot   1^{\shift}_{R^+_j}}
\end{align*}

In step $(**)$ we have used the definition of $n(\cl)$ and Eq.
\eqref{eq_cond_on_G}. Moreover, we have used
$$  \exp\bigl( i \tfrac{k+\cg}{2\pi}   \int_{\Sigma} \Tr\bigl(dA^{\orth}_{sing}(\cl) \cdot \tfrac{1}{2 \pi (k+\cg)}
 \sum_{j=1}^n  \bar{\alpha}_j \cdot   1^{\shift}_{R^+_j}\bigr)\bigr)
= \exp\bigl(- i \sum_j  \int_{l^j_{\Sigma}}  ( \bar{\alpha}_j, A^{\orth}_{sing}(\cl)) \bigr)
$$
 Step $(***)$ follows, informally by interchanging $\sum_{\cl} \cdots $
and $\int_{\ct} db \cdots$ and  then  using
 $$\sum_{x \in I } \exp(  2\pi i (k+\cg) ( x,b))
  = \sum_{b' \in \frac{1}{ k+\cg} I^*}  \delta(b-b') $$
 which is an informal version of the Poisson
summation formula (moreover, one has to take into account Remark
 \ref{rem4} and the relation $I^*=\Lambda$, cf. Remark \ref{rem1}).\par

The ``framing'' procedure mentioned above which has to be used for a
rigorous treatment can also be ``implemented'' in the heuristic
setting  we work with in the present paper. This amounts to replacing (by
hand) the expressions  $B(\sigma^{j}_m)$ appearing above
by\footnote{in a rigorous treatment where the Hida distributions
$\Psi_{\bar{\phi}_s}$, $s>0$, are used instead of the heuristic
integral functional $ \int \cdots \exp( i \tfrac{k+\cg}{2\pi}
\int_{\Sigma} \Tr(dA^{\orth}_c \cdot B)) (DA_c^{\orth} \otimes
   DB)$ a suitably regularized version of this linear combination
$\tfrac{1}{2} \bigl[ B(\bar{\phi}_s(\sigma^{j}_m)) +
B(\bar{\phi}_s^{-1}(\sigma^{j}_m))\bigr]$ appears naturally as a
result of the application of the polarization identity for
quadratic forms.}
 $\tfrac{1}{2} \bigl[ B(\bar{\phi}_s(\sigma^{j}_m)) + B(\bar{\phi}_s^{-1}(\sigma^{j}_m))\bigr]$.
Accordingly, one can expect that in the rigorous treatment
where $\WLO(L)$ is defined  and computed rigorously
 one has
\begin{equation} \label{eq_WLO_Endformel}
 \WLO(L) = C_1 \cdot \cS t_{CS}(L)
\end{equation}
where $C_1$ is a suitable constant independent of $L$ (see Eq.
\eqref{eq_def_C1} below) and where $\cS t_{CS}(L) $ is the
rigorous finite state sum (called  the ``Chern-Simons state sum of
$L$ in horizontal framing'' in the sequel) given by

\begin{multline} \label{eq_def_St}
\cS t_{CS}(L) :=    \sum_{\alpha_1, \alpha_2, \ldots, \alpha_n \in \Lambda}  \sum\nolimits_{b \in \frac{1 }{k+\cg} \Lambda}   \bigl(\prod_{j=1}^n m_{\gamma_j}(\alpha_j)\bigr)
   \biggl(  1_{(\ct_{reg})^{\Sigma}}(B) 1_{P}(B(\sigma_0))
   \det\nolimits_{reg}\bigl(1_{\ck}-\exp(\ad(B)_{| \ck})\bigr) \\
 \times  \prod_{j=1}^n    \exp( 2 \pi i
     ( \alpha_j, \sum\nolimits_{m} \eps^j_m  \tfrac{1}{2} \bigl[ B(\bar{\phi}_s(\sigma^{j}_m)) + B(\bar{\phi}_s^{-1}(\sigma^{j}_m))\bigr]) ) \biggr)_{| B=b +  \tfrac{1}{k+\cg} \sum_{j=1}^n  \alpha_j \cdot   1^{\shift}_{R^+_j} }
\end{multline}
where $s>0$ is chosen small enough\footnote{From Lemma \ref{lemma1} iii) below it follows that
 the  right-hand side of Eq. \eqref{eq_def_St} as a function of $s$ is stationary
as $s \to 0$, so it is clear what ``small enough'' means. Moreover,
Lemma \ref{lemma1} iii) below shows  that $\cS t_{CS}(L)$
 does not depend on the special choice of the horizontal framing $(\phi_s)_{s>0}$.}.

In the special case  $n=0$, i.e. the case where the link $L$ is ``empty'', it follows from
the heuristic definition of $\WLO(L)$ that we must have
$\WLO(L)=1$. From this and Eqs. \eqref{eq_WLO_Endformel},
\eqref{eq_def_St}, and \eqref{eq_detreg}  we can therefore
conclude
 \be \label{eq_def_C1}
C_1 = \biggl(\sum\nolimits_{b \in P \cap \frac{1 }{k+\cg} \Lambda
}
     \det\bigl(1_{\ck}-\exp(\ad(b))_{|\ck}\bigr)^{1-g}\biggr)^{-1}
\ee
where $g$ is the genus of $\Sigma$.
In Sec. \ref{sec5} below we will give  a somewhat more explicit
expression for $C_1$.

 \section{Equivalence of
  the Chern-Simons state sums and those in the shadow invariant}
\label{sec5}

 \begin{theorem} \label{theorem}
 Let $L$ be the colored link in $\Sigma \times S^1$
 which we have fixed above. Then
 \begin{equation}
   \cS t_{CS}(L) =  K^{2-2g} \cdot |X_L|
  \end{equation}
where $g$ is the genus of $\Sigma$ and
\begin{equation}K:= \prod_{\beta \in \cR_+}
  \bigl( 2\sin\bigl(  \tfrac{ \pi ( \beta,   \rho)}{k+\cg}\bigr) \bigr)
   \end{equation}
  \end{theorem}

  Before we prove Theorem \ref{theorem} we will first introduce some notation and
  then state and prove two lemmas.
  The proof of Theorem \ref{theorem}  will  be given after the proof of Lemma \ref{lemma2} below.\par

 For each  sequence $(\alpha_i)_{0 \le i \le n}$
of elements of $\Lambda$ we set
$$B_{(\alpha_i)_i}:=
\tfrac{1}{k+\cg} \bigl(  \alpha_0 +   \sum_{j=1}^n  \alpha_j \cdot
1^{\shift}_{R^+_j} \bigr).$$
 Then we can rewrite Eq.
\eqref{eq_def_St} as
\begin{multline} \label{eq_starting_formula}
\cS t_{CS}(L) =    \sum_{(\alpha_i)_i \in \Lambda^{n+1}}
  \bigl(\prod_{j=1}^n m_{\gamma_j}(\alpha_j)\bigr)    1_{P}(B_{(\alpha_i)_i}(\sigma_0)) 1_{(\ct_{reg})^{\Sigma}}(B_{(\alpha_i)_i})
  \det\nolimits_{reg}\bigl(1_{\ck}-\exp(\ad(B_{(\alpha_i)_i})_{| \ck})\bigr)  \\
 \times  \prod_{j=1}^n    \exp( 2\pi i
     ( \alpha_j, \sum\nolimits_{m} \eps^j_m  \tfrac{1}{2} \bigl[ B_{(\alpha_i)_i}(\bar{\phi}_s(\sigma^{j}_m)) + B_{(\alpha_i)_i}(\bar{\phi}_s^{-1}(\sigma^{j}_m))\bigr]) \rangle
  \end{multline}

 Each  $B_{(\alpha_i)_i}$ gives rise to
an ``area coloring''  $\varphi_{(\alpha_i)_i}: \{Y_0,Y_1, \ldots,
Y_{n}\} \to \Lambda $ given
 by
\begin{equation} \label{eq_def_varphi_alpha}
\varphi_{(\alpha_i)_i}(Y_t):=  (k+\cg)  B_{(\alpha_i)_i}(\sigma_{Y_t}) - \rho =
 \alpha_0 +   \sum_{j=1}^n  \alpha_j \cdot   1^{\shift}_{R^+_j}(\sigma_{Y_t})   - \rho
\end{equation}
where $\sigma_{Y_t}$ is an arbitrary point of $Y_t$. Note that
$\rho \in \Lambda$  so $\varphi_{(\alpha_i)_i}$ is
well-defined.\par

 \begin{lemma} \label{lemma1} For each $(\alpha_i)_{0 \le i \le n} \in \Lambda^{n+1}$ we have
 \begin{enumerate}
 \item[i)] $\alpha_j =   \varphi_{(\alpha_i)_i}(Y^+_j)    -  \varphi_{(\alpha_i)_i}(Y^-_j)$ for  $1 \le j \le n$
 \item[ii)] $\det\nolimits_{reg}\bigl(1_{\ck}-\exp(\ad(B_{(\alpha_i)_i})_{| \ck})\bigr)
=  K^{2-2g} \prod_Y(\dim(\varphi_{(\alpha_i)_i}(Y)))^{\chi(Y)}$
 \item[iii)] $ \prod_{j=1}^n    \exp( 2\pi i
     ( \alpha_j, \sum\nolimits_{m} \eps^j_m  \tfrac{1}{2} \bigl[ B_{(\alpha_i)_i}(\bar{\phi}_s(\sigma^{j}_m))
     + B_{(\alpha_i)_i}(\bar{\phi}_s^{-1}(\sigma^{j}_m))\bigr]) ) =
  \prod_{Y} \bigl(v_{\varphi_{(\alpha_i)_i}(Y)}\bigr)^{
\gleam(Y)}$
\end{enumerate}
  \end{lemma}

{\em Proof of i):} By Assumption \ref{assump1} the loops
$l^1_{\Sigma}, l^2_{\Sigma}, \ldots, l^n_{\Sigma}$
  do not intersect.
  This means that  for each $j'\le n$ with $j' \neq j$ the two faces
 $Y^+_j$ and $Y^-_j$ are either both ``inside'' $l^{j'}_{\Sigma}$ or both ``outside''
 $l^{j'}_{\Sigma}$. More precisely, we have
either $Y^+_j, Y^-_j \subset R^+_{j'}$ or $Y^+_j, Y^-_j \subset R^-_{j'}$.
 From Eq. \eqref{eq_def_varphi_alpha} we therefore obtain
$ \varphi_{(\alpha_i)_i}(Y^+_j)    -
\varphi_{(\alpha_i)_i}(Y^-_j)= \alpha_j \cdot
1^{\shift}_{R^+_j}(\sigma_{Y^+_j}) -
    \alpha_j \cdot 1^{\shift}_{R^+_j}(\sigma_{Y^-_j})=  \alpha_j \cdot 1_{R^+_j}(\sigma_{Y^+_j}) -
    \alpha_j \cdot 1_{R^+_j}(\sigma_{Y^-_j})= \alpha_j \cdot (1 -0) = \alpha_j$.\par

\medskip

 {\em Proof of ii):}   For every $b \in \ct$ we have
  \begin{multline} \label{eq_det=prod}
  \det(1_{\ck}-\exp(\ad(b))_{|\ck}) = \prod_{\beta \in \cR_+} (1-e^{2 \pi i \beta(b)})(1-e^{-2 \pi i \beta(b)})
 = \prod_{\beta \in \cR_+} \bigl[ - ( e^{2 \pi i \beta(b)/2} -e^{-2 \pi i \beta(b)/2})^2 \bigr] \\
 =  \prod_{\beta \in \cR_+} \bigl[ - ( e^{ \pi i ( \beta, b ) } -e^{-\pi i ( \beta, b )})^2 \bigr]
 = \prod_{\beta \in \cR_+} \bigl[ - (2i \sin( \pi ( \beta, b )))^2 \bigr]
 =  \prod_{\beta \in \cR_+} 4 \sin( \pi ( \beta, b ))^2
 \end{multline}

  Taking into account
  Eqs. \eqref{eq_detreg}, \eqref{eq_def_varphi_alpha}, and the relation $\sum_Y \chi(Y) = \chi(\Sigma) = 2-2g$
 we obtain
 \begin{multline}
   \det\nolimits_{reg}\bigl(1_{\ck}-\exp(\ad(B_{(\alpha_i)_i})_{| \ck})\bigr)
   =  \prod_{t=0}^{n} \prod_{\beta \in \cR_+}
    \bigl( \bigl(2 \sin( \pi ( \beta, B_{(\alpha_i)_i}(\sigma_{Y_t}))) \bigr)^2\bigr)^{\chi(Y_t)/2} \\
=  \prod_{t=0}^{n} \prod_{\beta \in \cR_+}  \bigl(2 \sin( \pi (
\beta, \tfrac{1}{k+\cg} \bigl( \varphi_{(\alpha_i)_i}(Y_t) + \rho
\bigr) ))\bigr)^{\chi(Y_t)} =   \prod_{Y} \prod_{\beta \in \cR_+}
\bigl(2 \sin(  \tfrac{\pi}{k+\cg} ( \beta,
\varphi_{(\alpha_i)_i}(Y) + \rho)
 \bigr)^{\chi(Y)}\\
=   K^{\chi(\Sigma)} \prod_{Y} \prod_{\beta \in \cR_+}
   \biggl( \frac{\sin(  \tfrac{\pi( \beta,  \varphi_{(\alpha_i)_i}(Y) + \rho)} {k+\cg})}{\sin(  \frac{ \pi ( \beta,   \rho)}{k+\cg})}
 \biggr)^{\chi(Y)} =  K^{2-2g}  \prod_Y(\mbox{dim}(\varphi_{(\alpha_i)_i}(Y)))^{\chi(Y)}
 \end{multline}

 \medskip

 {\em Proof of iii):}
 Recall that the framing $(\phi_s)_{s >0}$ was assumed to be horizontal.
Thus, for fixed $j$ and $m$, exactly one of the two points
$\bar{\phi}_s(\sigma^{j}_m)$ and $\bar{\phi}_s^{-1}(\sigma^{j}_m)$
will lie in $Y^{+}_j$ and the other one in $Y^{-}_j$ and  we have
for sufficiently small $s>0$
\begin{equation}
B_{(\alpha_i)_i}(\bar{\phi}_s(\sigma^{j}_m)) +
B_{(\alpha_i)_i}(\bar{\phi}_s^{-1}(\sigma^{j}_m))
  = \tfrac{1}{k+\cg} ( \varphi_{(\alpha_i)_i}(Y^{+}_j) + \varphi_{(\alpha_i)_i}(Y^{-}_j) + 2\rho)
\end{equation}

Let us set
\begin{equation} \label{eq_def_eps_j} \eps_j :=  \sum_{m \le n^j}    \eps^j_{m} = \wind(l_{S^1}^j)
\end{equation}
Then, taking into account part i) of the Lemma we get (for small $s>0$)
\begin{align*}
&  \prod_{j}  \exp( 2\pi i  ( \alpha_j, \sum\nolimits_{m} \eps^j_m  \tfrac{1}{2} \bigl[ B_{(\alpha_i)_i}(\bar{\phi}_s(\sigma^{j}_m)) + B_{(\alpha_i)_i}(\bar{\phi}_s^{-1}(\sigma^{j}_m))\bigr]) )\\
& =   \prod_{j}
\exp( 2 \pi i \bigl( \sum\nolimits_{m} \eps^j_m \bigr) \tfrac{1}{2}
\tfrac{1}{k+\cg}  \bigl( \varphi_{(\alpha_i)_i}(Y^+_j)    -  \varphi_{(\alpha_i)_i}(Y^-_j),   \varphi_{(\alpha_i)_i}(Y^+_j) + \varphi_{(\alpha_i)_i}(Y^{-}_j) + 2\rho        \bigr)\\
 & =   \prod_{j} \exp\biggl( \tfrac{\pi i}{k+\cg} \eps_j
   \biggl[ ( \varphi_{(\alpha_i)_i}(Y^+_j), \varphi_{(\alpha_i)_i}(Y^+_j) + 2\rho) -
  ( \varphi_{(\alpha_i)_i}(Y^-_j), \varphi_{(\alpha_i)_i}(Y^-_j) + 2\rho) \biggr] \biggr) \\
  &   =   \prod_{j} \exp\bigl(   \tfrac{\pi i}{k+\cg} \eps_j  \bigl[ \sgn(Y^+_j; l^{j}_{\Sigma})
  C_2(\varphi_{(\alpha_i)_i}(Y^+_j)) + \sgn(Y^{-}_j; l^{j}_{\Sigma})
   C_2(\varphi_{(\alpha_i)_i}(Y^-_j))  \bigr] \bigr) \\
     &  =  \prod_{Y} \exp\biggl(   \tfrac{\pi i}{k+\cg}  \biggl(\sum\nolimits_{j \text{ with } \arc(l^j_{\Sigma}) \subset \partial Y}  \eps_{j}   \sgn(Y;l^j_{\Sigma})\biggr) C_2(\varphi_{(\alpha_i)_i}( Y)\bigr) \biggr) \\
&  \overset{(*)}{=}  \prod_{Y}  \exp\biggl(   \tfrac{\pi i}{k+c_g}
\ \gleam(Y) \cdot C_2(\varphi_{(\alpha_i)_i}( Y)\bigr)  \biggr) =
\biggl( \prod_{Y} \bigl(v_{\varphi_{(\alpha_i)_i}(Y)}\bigr)^{
\gleam(Y)} \biggr) \cdot \bigl( \prod_{Y} e^{\frac{\pi i c}{12}
\gleam(Y)} \bigr) \\
&  \overset{(**)}{=} \biggl( \prod_{Y}
\bigl(v_{\varphi_{(\alpha_i)_i}(Y)}\bigr)^{ \gleam(Y)} \biggr)
\end{align*}
Here step $(*)$ follows from Eq. \eqref{eq_formel_gleam} and
 Eq. \eqref{eq_def_eps_j}.
 Moreover, also step $(**)$
follows from  Eq. \eqref{eq_formel_gleam} which clearly implies
$\sum_Y \gleam(Y) = 0$.

\medskip

  Recall that ${\mathrm col}(X_L)$ denotes the set of
mappings  $\{Y_0,Y_1, \ldots, Y_{n}\} \to  \Lambda_+^k$. In the
sequel let  ${\mathrm col'}(X_L)$  denote the set of mappings
$\{Y_0,Y_1, \ldots, Y_{n}\} \to \Lambda  \cap ((k+\cg) \ct_{reg} -
\rho) $ and let $(\cW_k)^{\{Y_0,Y_2, \ldots, Y_{n}\}}$, or simply,
$(\cW_k)^{n+1}$ denote  the space of functions from $\{Y_0,Y_1,
\ldots, Y_{n}\}$ with values in $\cW_k$.

 \begin{lemma}  \label{lemma2}
  The mappings
 $$\Phi: \{ (\alpha_i)_{0 \le i \le n} \in \Lambda^{n+1} \mid  1_{P^{\Sigma}}(B_{(\alpha_i)_i}) \neq 0 \}
   \ni (\alpha_i)_{0 \le i \le n} \mapsto \varphi_{(\alpha_i)_i} \in {\mathrm col}(X_L)$$
   $$\Phi':  \{ (\alpha_i)_{0 \le i \le n} \in \Lambda^{n+1} \mid  1_{(\ct_{reg})^{\Sigma}}(B_{(\alpha_i)_i}) \neq 0 \} \ni (\alpha_i)_{0 \le i \le n} \mapsto \varphi_{(\alpha_i)_i}
\in {\mathrm col'}(X_L)$$
  are  well-defined bijections and we have
  \begin{equation} \label{eq_colX_colX'}
 col'(X_L) =  \{ \underline{\tau}  \cdot \varphi \mid  \varphi \in col(X_L),
    \underline{\tau}  \in (\cW_k)^{n+1} \}
\end{equation}
where $ \underline{\tau}  \cdot \varphi \in  (\cW_k)^{n+1}$ is  given by
 $ (\underline{\tau}  \cdot \varphi)(Y) = \underline{\tau}(Y) \cdot \varphi(Y)$ for all $Y \in \{Y_0,Y_1, \ldots, Y_{n}\}$.
  \end{lemma}

 \begin{proof}  \begin{enumerate}
 \item {\rm \em $\Phi'$ is  well-defined and surjective:}
 Clearly, we have
 $\{ \varphi_{(\alpha_i)_i}    \mid (\alpha_i)_{0 \le i \le n} \in \Lambda^{n+1} \} = \Lambda ^{\{Y_0,Y_1, \ldots, Y_{n}\} }$.
On the other hand, for fixed  $(\alpha_i)_{0 \le i \le n} \in \Lambda^{n+1}$
  the relation  $1_{(\ct_{reg})^{\Sigma}}(B_{(\alpha_i)_i}) \neq 0$
  is equivalent to
  $\Image(B_{(\alpha_i)_i}) = \Image(\tfrac{1}{k+\cg} (\varphi_{(\alpha_i)_i} + \rho))  \subset \ct_{reg}$
   which is equivalent to
    $\Image(\varphi_{(\alpha_i)_i}) \subset (k+\cg) \ct_{reg} - \rho$. The assertion now follows.

 \item {\rm \em $\Phi$ is well-defined and surjective:} For fixed $(\alpha_i)_{0 \le i \le n} \in \Lambda^{n+1}$
   the relation  $1_{P^{\Sigma}}(B_{(\alpha_i)_i}) \neq 0$
   is equivalent to $\Image(B_{(\alpha_i)_i}) = \Image(\tfrac{1}{k+\cg} (\varphi_{(\alpha_i)_i} + \rho)) \subset P$
   which is equivalent to $\Image(\varphi_{(\alpha_i)_i}) \subset  (k+\cg) P - \rho$.
   Thus the assertion follows if we can show that
\begin{equation} \label{eq_P<->Lambda+k}
\Lambda  \cap ((k+\cg) P - \rho) =  \Lambda^{k}_+
\end{equation}
 In order to prove this equation note that
 $P = \CW \cap  \{\lambda  \in \ct \mid (\lambda,\theta) < 1 \}$
so we have
\begin{align*}
& \Lambda  \cap ((k+\cg) P - \rho)\\
& =  \Lambda    \cap  \bigl( \CW \cap  \{\lambda  \in \ct \mid (\lambda,\theta) < k+\cg \} - \rho \bigr) \\
& =  \Lambda \cap (\CW - \rho) \cap   \{\lambda  \in \ct \mid (\lambda + \rho ,\theta) < k+\cg \}\\
 &  \overset{(*)}{=}  \Lambda \cap \overline{\CW}  \cap   \{\lambda  \in \ct \mid (\lambda + \rho ,\theta) < k+\cg \}\\
&  = \Lambda_+  \cap  \{\lambda  \in \ct \mid(\lambda   ,\theta) < k+\cg - (\rho ,\theta) \}\\
&  \overset{(**)}{=}  \{\lambda  \in \Lambda_+ \mid (\lambda   ,\theta) < k+1  \} \\
& \overset{(***)}{=} \{\lambda  \in \Lambda_+ \mid (\lambda   ,\theta) \le k  \} = \Lambda^{k}_+
\end{align*}
Here  step $(*)$ follows because for each $\alpha \in  \Lambda$,
    $\alpha + \rho $ is in the open Weyl chamber $\CW$ iff $\alpha$
    is in the closure $\overline{\CW}$, i.e. we have
    $\Lambda  \cap (\CW - \rho) = \Lambda  \cap \overline{\CW} = \Lambda_+$
  (cf. the last remark   in Sec. V.4 in \cite{Br_tD}).
Step $(**)$ follows from  $\cg =1+ (\theta,\rho)$ and step $(***)$ from $(\lambda ,\theta) \in \bZ$ for each $\lambda \in \Lambda$.

 \item {\rm \em Formula \eqref{eq_colX_colX'} holds:}  This follows  from the fact that both the mapping
 $\cW_{\aff} \times P \ni (\tau,b) \mapsto \tau \cdot b \in \ct_{reg}$
and the mapping  $i:\cW_{\aff} \to \cW_k$ in Subsec. \ref{subsec2.2}
are bijections.

 \item {\rm \em $\Phi$ and $\Phi'$ are injective:}
 Let
   $(\alpha''_i)_i, (\alpha'_i)_i \in \{ (\alpha_i)_{0 \le i \le n}
    \in \Lambda^{n+1} \mid  1_{(\ct_{reg})^{\Sigma}}(B_{(\alpha_i)_i})\neq 0 \}$
    such that $\varphi_{(\alpha''_i)_i} = \varphi_{(\alpha'_i)_i}$.
    From Lemma \ref{lemma1} i) it then follows immediately
    that $\alpha''_i = \alpha'_i$ for $i \in \{1,2,\ldots,n\}$.
    Moreover, from Eq. \eqref{eq_def_varphi_alpha} and Eq. \eqref{eq_1_shift}
 we get $\alpha''_0 =\varphi_{(\alpha''_i)_i}(Y_{\sigma_0}) + \rho  =  \varphi_{(\alpha'_i)_i}(Y_{\sigma_0}) + \rho = \alpha'_0$
 where  $Y_{\sigma_0}$ denotes the face which contains the point $\sigma_0$.
 \end{enumerate}

  \end{proof}

 \medskip

\noindent {\bf Proof of Theorem \ref{theorem}:}  Applying  Lemma \ref{lemma1} to Eq. \eqref{eq_starting_formula} we obtain
\begin{multline}
 \cS t_{CS}(L)
=   K^{2-2g}  \sum_{(\alpha_i)_i \in \Lambda^{n+1}}
1_{P}(B_{(\alpha_i)_i}(\sigma_0))
 1_{(\ct_{reg})^{\Sigma}}(B_{(\alpha_i)_i})
  \bigl(\prod_{j=1}^n  m_{\gamma_j}(\varphi_{(\alpha_i)_i}(Y^+_j)    -  \varphi_{(\alpha_i)_i}(Y^-_j))\bigr)  \\ \times \prod_Y(\mbox{dim}(\varphi_{(\alpha_i)_i}(Y)))^{\chi(Y)}
    \biggl( \prod_{Y} (v_{\varphi_{(\alpha_i)_i}}(Y))^{ \gleam(Y)} \biggr)
  \end{multline}

 Without loss of generality we can assume that $\sigma_0 \in Y_0$.
 Then  we obtain from  Lemma \ref{lemma2}
\begin{multline} \label{eq70}
 \cS t_{CS}(L)
 =  K^{2-2g}  \sum_{\varphi \in col'(X_L)}  1_{P}((k+\cg) \cdot (\varphi(Y_0) + \rho))
  \bigl(\prod_{j=1}^n  m_{\gamma_j}(\varphi(Y^+_j)    -  \varphi(Y^-_j))\bigr) \\
  \times   \prod_Y(\dim(\varphi(Y)))^{\chi(Y)}     \biggl( \prod_{Y} (v_{\varphi(Y)})^{ \gleam(Y)} \biggr)
\end{multline}

Now observe that   for all $\tau \in \cW_k$, $b \in  \Lambda \cap ( (k+\cg) \ct_{reg} - \rho)$ we have
\begin{align}
  v_{\tau \cdot b} & = v_b\\
  \dim(\tau \cdot b) & = \sgn(\tau) \dim(b)
\end{align}

 Moreover, $1_{P}((k+\cg) ( \tau \cdot \varphi(Y_0) + \rho)) = 1_{\tau=1}$
 for $\varphi \in col(X_L)$, $\tau \in \cW_k$.
 Thus we obtain from Eq.   \eqref{eq70} and Eq. \eqref{eq_colX_colX'} in Lemma \ref{lemma2}
  \begin{align} \label{eq_change_to_tau_tilde}
 \cS t_{CS}(L)
&  =  K^{2-2g}   \sum_{\varphi \in col(X_L)}
\sum_{\underline{\tau} \in (\cW_k)^{n+1}  }
1_{\tau_0=1} \bigl(\prod_{j=1}^n  m_{\gamma_j}(\underline{\tau}(Y^+_j) \cdot \varphi(Y^+_j)    -  \underline{\tau}(Y^-_j) \cdot \varphi(Y^-_j)\bigr) \nonumber \\
& \quad \quad \times \prod_{Y} \bigl(\sgn(\underline{\tau}(Y))\bigr)^{\chi(Y)}   \prod_Y(\dim(\varphi(Y)))^{\chi(Y)}     \biggl( \prod_{Y} (v_{\varphi(Y)})^{ \gleam(Y)} \biggr) \nonumber \\
& =   K^{2-2g} \sum_{\varphi \in col(X_L)} \sum_{\tau_0, \tau_1,
\ldots \tau_{n} \in \cW_k} 1_{\tau_0=1}
 \prod_{t=0}^{n} \bigl(\sgn(\tau_t)\bigr)^{\chi(Y_t)}
  \bigl(\prod_{j=1}^n  m_{\gamma_j}( \tau_{t(j,+)} \cdot \varphi(Y^+_j)    -  \tau_{t(j,-)} \cdot \varphi(Y^-_j)\bigr) \nonumber \\
& \quad \quad \times   \prod_Y(\dim(\varphi(Y)))^{\chi(Y)}     \biggl( \prod_{Y} (v_{\varphi(Y)})^{ \gleam(Y)} \biggr)
\end{align}
 where $t(j,+)$ resp. $t(j,-)$ is the unique index $t \in \{0,1,2,\ldots,n\}$
  such that  $Y_t = Y^+_j$ resp.  $Y_t = Y^-_j$ holds.
  Each  $m_{\gamma_j}$ is invariant under the (classical) Weyl group $\cW$.
  From  \eqref{eq_def_Wq} and \eqref{eq_Wq_Waff} and the fact that each $\tau \in \cW_{\aff}$
  can be written as the product of a translation and an element of $\cW$
  it easily follows  that
\begin{equation}
 m_{\gamma_j}(\tau_{t(j,+)} \cdot \varphi(Y^+_j)    -  \tau_{t(j,-)} \cdot \varphi(Y^-_j)\bigr)
= m_{\gamma_j}(\varphi(Y^+_j)    - \tau_{t(j,+)}^{-1} \cdot \tau_{t(j,-)} \cdot \varphi(Y^-_j)\bigr)
\end{equation}

Accordingly, let us  set $\tilde{\tau}_j := \tau_{t(j,+)}^{-1}
\cdot \tau_{t(j,-)}$. Clearly, we have
\begin{equation} \label{eq_signs}
 \prod_{t=0}^n \bigl(\sgn(\tau_t)\bigr)^{\chi(Y_t)} \overset{(*)}{=}
 \prod_{t=0}^n \bigl(\sgn(\tau_t)\bigr)^{\# \{ j \le n \mid \arc(l^j_{\Sigma}) \subset \partial Y_t\}}
 = \prod_{j=1}^n \sgn(\tau_{t(j,+)}) \sgn(\tau_{t(j,-)})
  =  \prod_{j=1}^n \sgn(\tilde{\tau}_j)
 \end{equation}
(here step $(*)$ follows from $\chi(Y_t) = 2 -\# \{ j \le n \mid
\arc(l^j_{\Sigma}) \subset \partial Y_t\}$). On the other hand the
expressions $\sum_{\tilde{\tau}_j}  \sgn(\tilde{\tau}_j)
\bigl(\prod_{j=1}^n  m_{\gamma_j}(  \varphi(Y^+_j)    -
\tilde{\tau}_j \cdot \varphi(Y^-_j))\bigr)$,  $j \le n$, are
exactly of the form of the expression on the right-hand side of
formula \eqref{eq_quantum_racah} so we have
 $$\sum_{\tilde{\tau}_1, \tilde{\tau}_2, \ldots \tilde{\tau}_n \in \cW_k}
\prod_{j=1}^n \sgn(\tilde{\tau}_j)   \bigl(\prod_{j=1}^n  m_{\gamma_j}(  \varphi(Y^+_j)    - \tilde{\tau}_j \cdot \varphi(Y^-_j)\bigr) =
 \prod_{j=1}^n N_{ \gamma_j  \varphi(Y^+_j)}^{ \varphi(Y^-_j)}
 $$
 Combining this with Eqs. \eqref{eq_change_to_tau_tilde}--\eqref{eq_signs}
 we obtain
\begin{align*}
  \cS t_{CS}(L)
 & =  K^{2-2g} \sum_{\varphi \in col(X_L)}   \bigl(\prod_{j=1}^n N_{ \gamma_j  \varphi(Y^+_j)}^{ \varphi(Y^-_j)} \bigr)   \prod_Y(\dim(\varphi(Y)))^{\chi(Y)}
    \biggl( \prod_{Y} (v_{\varphi(Y)})^{ \gleam(Y)} \biggr)\\
 & =  K^{2-2g} \sum_{\vf\in\sm{col}(X_L)} |X_L|_1^\vf
|X_L|_2^\vf |X_L|_3^\vf \\
& =  K^{2-2g}  |X_L|  \quad \quad \text{\bf q.e.d.}
\end{align*}

From  Theorem \ref{theorem} and Eq.  \eqref{eq_WLO_Endformel}
above we can  conclude that $\WLO(L)$ coincides with $|X_L|$ up to
a multiplicative constant (independent of $L$). We can easily
determine this multiplicative constant explicitly.
 According to Eqs.
\eqref{eq_def_C1}, \eqref{eq_det=prod}, and  Eq.
\eqref{eq_def_dim} we have (cf. Eq. \eqref{eq_P<->Lambda+k} and
Example \ref{Ex3} above)
\begin{align} \label{eq_expl_C1}
 C_1 & = \frac{1}{ \sum\nolimits_{\lambda \in \Lambda^k_+ }
     ( K  \dim(\lambda))^{2-2g} }
 = \frac{1}{  K^{2-2g}}  \frac{1}{|X_{\emptyset}|}
\end{align}
so from   Eq. \eqref{eq_WLO_Endformel} and
 Theorem \ref{theorem}
 we finally  obtain
\begin{equation} \label{eq_WLO=shadowinv}  \WLO(L) =  \frac{ |X_L|}{|X_{\emptyset}|}
\end{equation}
This agrees exactly with the formula appearing at the end of
Subsec. \ref{subsec4.1} above.

\section{A  path integral derivation of the quantum Racah formula}
\label{sec7}

In \cite{BlTh1}  $\WLO(L)$ was evaluated in the torus gauge
approach in the special case  where the link $L$ consists
exclusively of 3 vertical loops with colors $\lambda, \mu, \nu \in
\Lambda^k_+$  (cf. Remark \ref{rm_BlTh} below).
 The result of this evaluation is  the expression on the right-hand side of Eq.
\eqref{rhs_fusion_rules} above. As we showed in Secs. \ref{sec4}--\ref{sec5}
 when evaluating the WLOs of  loops without
double points in the torus gauge approach the expressions on the
right-hand side of  Eq.
 \eqref{eq_quantum_racah} arise
naturally. In other words: both the left-hand side and the
right-hand side of Eq. \eqref{eq_quantum_racah} appear naturally
in the torus gauge approach when computing the WLOs of suitable
links.  One can
therefore hope to obtain a path integral derivation of Eq.
\eqref{eq_quantum_racah} by considering links that contain both
vertical loops and  loops without double points. Accordingly, let
us now generalize some of the results obtained in Secs. \ref{sec4}
and \ref{sec5}
 to this more general situation where the (colored) link  $L= ((l_1, l_2, \ldots,
l_N),(\gamma_1,\gamma_2,\ldots,\gamma_N))$ is allowed to contain
vertical loops. More precisely, we assume
 that  the sub link $(l_1,l_2, \ldots, l_n)$, $n \le N$, is admissible
 and each loop $l_k$ for $k \in \{n+1, \ldots, N\}$
 is a ``vertical'' loop ``above'' the point $\sigma_k \in \Sigma$,
 i.e. $l^k_{\Sigma}$ is a constant mapping taking only the value
 $\sigma_k$.
 For  each of the vertical loops $l_k$, $k \in \{n+1, \ldots, N\}$,
 we will use a   ``canonical'' framing, i.e. a framing  which fulfills the following condition:
  if the framing is represented\footnote{alternatively, if we represent the framing in terms
  of  another loop $l'_k$ which is ``sufficiently close'' to $l_k$ (cf. e.g. Sec. 2.1 and Fig. 3b and Fig. 3c in \cite{Wi})  then a canonical framing is one where (not only $l_k$ but also) $l'_k$
is vertical loop} in terms of a  vector field $X$ on $arc(l_k)$,
 then the projection of each  vector $X_{l_k(s)}$, $s \in [0,1]$,
 onto $T_{\sigma_k} \Sigma$ coincides with some fixed vector $v \in T_{\sigma_k} \Sigma$. \par

  Finally, we will assume for simplicity that $\wind(l^k_{S^1})=1$
for each $k \in \{n+1, \ldots, N\}$.\par
 Then, using similar
arguments as in Subsec. \ref{subsec4.4} we can again derive Eq.
\eqref{eq_WLO_Endformel} where $\cS t_{CS}(L)$ is now given by
\begin{multline} \label{eq_def_St_gen}
 \cS t_{CS}(L):=  \sum_{\alpha_1, \alpha_2, \ldots, \alpha_n \in \Lambda}  \sum\nolimits_{b \in \frac{1 }{k+\cg} \Lambda}   \bigl(\prod_{j=1}^n m_{\gamma_j}(\alpha_j)\bigr)
   \biggl(  1_{(\ct_{reg})^{\Sigma}}(B) 1_{P}(B(\sigma_0))
   \det\nolimits_{reg}\bigl(1_{\ck}-\exp(\ad(B)_{| \ck})\bigr) \\
  \times \bigl(\prod_{k=n+1}^{N} \chi_{\gamma_k}(\exp(B(\sigma_k)))\bigr)\\
 \times  \prod_{j=1}^n    \exp( 2 \pi i
     ( \alpha_j, \sum\nolimits_{k} \eps^j_k  \tfrac{1}{2} \bigl[ B(\bar{\phi}_s(\sigma^{j}_k)) + B(\bar{\phi}_s^{-1}(\sigma^{j}_k))\bigr]) ) \biggr)_{| B=b +  \tfrac{1}{k+\cg} \sum_{j=1}^n
      \alpha_j \cdot   1^{\shift}_{R^+_j} }
\end{multline}
for sufficiently small $s>0$. Recall that $
\chi_{\gamma_k}$ is the character associated to the dominant
weight $\gamma_k$.\par
Also the
computations in the proof of  Theorem \ref{theorem} can be
generalized in a straightforward way.
 One obtains
 \begin{multline} \label{eq_fast_fertig_mod} \cS t_{CS}(L)
=  K^{2-2g} \sum_{\varphi \in col(X_L)} \bigl(\prod_{j=1}^n M_{
\gamma_j  \varphi(Y^+_j)}^{ \varphi(Y^-_j)} \bigr)
\bigl(\prod_{k=n+1}^N \chi_{\gamma_k}(\exp(\tfrac{1}{k+\cg}
(\varphi(Y_{\sigma_k}) +
\rho) )) \bigr) \\
\times  \prod_Y(\dim(\varphi(Y)))^{\chi(Y)}
    \biggl( \prod_{Y} (v_{\varphi(Y)})^{ \gleam(Y)} \biggr)
 \end{multline}
 where $Y_{\sigma_k}$, $k \in \{n+1, \ldots, N\}$, denotes the face in which $\sigma_k$ lies
 and where we have set
\begin{equation}
M_{ \gamma \alpha}^{\beta} := \sum_{\tau \in \cW_{k}} \sgn(\tau)
m_{\gamma}(\alpha-\tau(\beta)) \end{equation}
 (According to Eq.
\eqref{eq_quantum_racah} we have $M_{ \gamma \alpha}^{\beta} = N_{
\gamma \alpha}^{\beta}$ so we could  replace $M_{ \gamma
\alpha}^{\beta}$ by $N_{ \gamma \alpha}^{\beta}$ above.
  But as  we want to give a path
integral derivation of Eq. \eqref{eq_quantum_racah}
 which is based on Eq. \eqref{eq_fast_fertig_mod} we avoid this
 replacement here.) Now observe that
\begin{equation} \label{eq_char_Smatrix}
\chi_{\mu}(\exp(\tfrac{1}{k+\cg} (\lambda + \rho) )) =
\frac{S_{\mu \lambda }}{S_{0 \lambda}}
\end{equation}
Eq. \eqref{eq_char_Smatrix} follows from the definition of the
S-matrix in Subsec. \ref{subsec2.2}
 if one takes into account  Weyl's character formula.
  Combining Eqs.  \eqref{eq_WLO_Endformel},  \eqref{eq_expl_C1}
\eqref{eq_fast_fertig_mod}, and \eqref{eq_char_Smatrix}
 we finally obtain
\begin{multline} \label{eq_WLO_end_gen}
\WLO(L) \\
= \frac{1}{|X_{\emptyset}|} \sum_{\varphi \in col(X_L)}
\bigl(\prod_{j=1}^n M_{ \gamma_j  \varphi(Y^+_j)}^{
\varphi(Y^-_j)} \bigr)  \bigl(\prod_{k=n+1}^N \frac{S_{\gamma_k
\varphi(Y_{\sigma_k}) }}{S_{0 \varphi(Y_{\sigma_k})}}\bigr)
\prod_Y(\dim(\varphi(Y)))^{\chi(Y)}
    \biggl( \prod_{Y} (v_{\varphi(Y)})^{ \gleam(Y)} \biggr)
\end{multline}

In the special case $n=0$, i.e. in the case where there are only vertical loops,
there is only one face $Y_0=\Sigma$ and we have $\gleam(Y_0)=0$,
$\chi(Y_0)= \chi(\Sigma)=2-2g$  so
 Eq. \eqref{eq_WLO_end_gen}
then reduces to
\begin{equation} \label{eq_WLO_end_vert}
\WLO(L)  = \frac{1}{|X_{\emptyset}|} \sum_{\lambda \in
\Lambda^k_+}
 \bigl(\prod_{k=1}^N \frac{S_{\gamma_k \lambda }}{S_{0 \lambda}}\bigr) \dim(\lambda)^{2-2g}
\end{equation}

\begin{remark} \rm \label{rm_BlTh}
In the special case $G=SU(2)$ the last equation is
 equivalent   to  formula (7.27) in \cite{BlTh1}.
We remark that for vertical loops the inner integral in Eq.
\eqref{eq_WLO_0} is trivial, so for the derivation of Eq.
\eqref{eq_WLO_end_vert}  one does not need the general formula
\eqref{eq_WLO_0} but can work with the simpler formulas appearing
in \cite{BlTh1}, cf. equations (7.1) and (7.24) in \cite{BlTh1}.
In the special case where $N=3$ and $\Sigma=S^2$, i.e. $g=0$
we get from Eq. \eqref{eq_WLO_end_vert}
(setting $\lambda:=\gamma_1$, $\mu:=\gamma_2$, $\nu:= \gamma_3$)
\begin{equation} \label{eq_WLO_end_3vert} \WLO(L)
= \frac{1}{|X_{\emptyset}|}  \sum_{\lambda_0}
  \frac{S_{\lambda \lambda_0}}{S_{0 \lambda_0}}   \frac{S_{\mu \lambda_0}}{S_{0 \lambda_0}}
   \frac{S_{\nu \lambda_0}}{S_{0 \lambda_0}}
 \dim(\lambda_0)^{2} = \frac{N_{\l\m\n}}{ \sum_{\alpha} S_{0
 \alpha}^2}= N_{\l\m\n}
 \end{equation}
 (here we have used $\sum_{\alpha} S_{0
 \alpha}^2 = (S \cdot S^T)_{00} =  (S^2)_{00} = C_{00} = 1$).
By combining Eq. \eqref{eq_WLO_end_3vert}  with Eq. (4.36) in
\cite{Wi} one obtains\footnote{Note  that, strictly speaking, this
is not quite a ``path integral derivation'' of the Verlinde formula
since the derivation of the Eq. (4.36) in \cite{Wi} is not based
solely on the CS path integral. In fact, since the numbers
$\hat{N}^{i}_{jk}$ are defined abstractly, a genuine
 path integral
derivation of the Verlinde formula \eqref{eq_Verlinde} can not be
expected.} the Verlinde formula \eqref{eq_Verlinde}, cf. Sec. 7.6 in
\cite{BlTh1}. (Observe that the expression $N_{ijk}$ in Eq. (4.36)
in \cite{Wi} does not correspond to $N_{ijk}=N^{i^*}_{jk} $
 in our notation but to $\hat{N}^{i^*}_{jk}$, cf. Sec. \ref{subsec2.2}).
\end{remark}

\begin{figure}[h]
\begin{center}
\includegraphics[height=1in,width=3in]{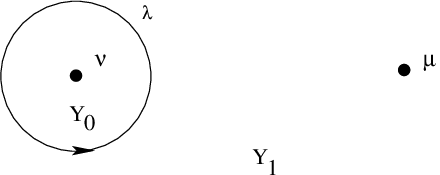}
\caption{} \label{smallcircles2}
\end{center}
\end{figure}

In order to obtain a path integral derivation of Eq.
\eqref{eq_quantum_racah} let us now restrict ourselves to the special case $\Sigma = S^2$
and consider a  link  $L$ in $M=\Sigma \times S^1 = S^2 \times S^1$
 which consists of 2 vertical loops $l_2$, $l_3$
over the point $\sigma_2$ resp. $\sigma_3$ with colors $\m$ and
$\n$ and one non-vertical  loop $l_1$ with color $\l$. We assume
that $\wind(l^i_{S^1})=1$ for all $i = 1,2,3$ and that
 $l^1_{\Sigma}$ is a Jordan loop (i.e. a simple loop without crossings).
 Moreover, we assume that $\sigma_2, \sigma_3$
are on different sides of $l^1_{\Sigma}$, i.e. that  the loop
projections $l^1_{\Sigma}, l^2_{\Sigma}, l^3_{\Sigma}$ look as in Fig. \ref{smallcircles2}

We will now evaluate $\WLO(L)$ in two different ways. By comparing
these two different evaluations of $\WLO(L)$ with each other we
will then obtain a system of linear equations for the ``unknowns''
$M_{ \gamma \alpha}^{\beta}$, $\a, \b, \g \in \Lambda^k_+$.
 Later we will show that the
unique solution of this system of linear equations is $M_{ \gamma
\alpha}^{\beta} =N_{ \gamma \alpha}^{\beta}$, $\a, \b, \g \in
\Lambda^k_+$,  which is nothing but formula
\eqref{eq_quantum_racah}.\\

\noindent {\bf 1. Evaluation:} Let us apply  Eq.
\eqref{eq_WLO_end_gen} to the link $L$. Observe that we have to
take $n=1$, $N=3$ and $(\gamma_1,\gamma_2,\gamma_3)= ( \l, \m,
\n)$.
 We obtain
\begin{equation} \label{eq_our_0}
\WLO(L) = \frac{1}{|X_{\emptyset}|}  \sum_{\varphi \in col(X_L)}
 M_{ \lambda  \varphi(Y^+_1)}^{ \varphi(Y^-_1)}
  \bigl(\prod_{k=2}^3
\frac{S_{\gamma_k \varphi(Y_{\sigma_k})
}}{S_{0\varphi(Y_{\sigma_k})}}\bigr)
 \prod_{t=0}^1 (\dim(\varphi(Y_t)))^{\chi(Y_t)}
    \biggl( \prod_{t=0}^1  (v_{\varphi(Y_t)})^{ \gleam(Y_t)} \biggr)
\end{equation}
Note that in this situation there are only two faces namely
 $Y_0:=  Y^+_1  $ and $Y_1: = Y^-_1$
 and we have $\sigma_2 \in Y_0$ and $\sigma_3 \in Y_1$
 (cf. Fig. \ref{smallcircles2}, note that the point $\sigma_2$  is labelled by the letter
$\nu$ and $\sigma_3$ by the letter $\mu$).
  Moreover, $\gleam(Y_0)=1$,  $\gleam(Y_1)=-1$, $\chi(Y_1)=1$,
$\chi(Y_0)=2-2g - 1=1$ (as $\Sigma=S^2$, so $g=0$). Using the
variables $\lambda_0:= \varphi(Y_0)$ and $\lambda_1:=
\varphi(Y_1)$ we now obtain
\begin{align} \label{eq_our}
 \WLO(L)   & =
\frac{1}{| X_{\emptyset}|} \sum_{\lambda_0,\lambda_1}
 M_{ \lambda  \lambda_0}^{\lambda_1} \
 \frac{S_{\nu \lambda_0}}{S_{0 \lambda_0}} \ \frac{S_{\mu \lambda_1}}{S_{0 \lambda_1}}
 \dim(\lambda_0) \dim(\lambda_1)
T_{\lambda_0 \lambda_0}  T_{\lambda_1 \lambda_1}^{-1} \nonumber\\
& = \frac{1}{| X_{\emptyset}|} \frac{1}{S_{00}^2}
\sum_{\lambda_0,\lambda_1}
 M_{ \lambda  \lambda_0}^{\lambda_1} \
S_{\nu \lambda_0} S_{\mu \lambda_1}  T_{\lambda_0 \lambda_0}
T_{\lambda_1 \lambda_1}^{-1}
\end{align}

\noindent {\bf 2. Evaluation:} Another explicit evaluation of $\WLO(L)$
can be obtained by considering, as an auxiliary object, the  colored link
$\hat{L}= ((\hat{l}_1, \hat{l}_2,
\hat{l}_3),(\lambda,\mu,\nu))$ in $M=S^2 \times S^1$
 where each $\hat{l}_j$, $j \in \{1,2,3\}$, is a vertical loop over the
point $\sigma_j$ with  $\wind(\hat{l}^j_{S^1}) = 1$, cf.
Fig. \ref{3points} below.
\begin{figure}[h]
\begin{center}
\includegraphics[height=.13in,width=2in]
{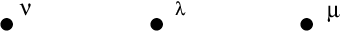} \caption{} \label{3points}
\end{center}
\end{figure}
  From  Eq. \eqref{eq_WLO_end_vert} (or Eq. \eqref{eq_WLO_end_3vert}) we
obtain $\WLO(\hat{L})  =  \frac{1}{|X_{\emptyset}|}
\frac{1}{S_{00}^2} N_{\l\m\n}$.\par

Let us now deform $\hat{L}$ using an orientation preserving
 diffeomorphism $\phi: S^2 \times S^1 \to S^2 \times S^1$ of the form
$\phi(\sigma,t) = \bigl(\theta_t(\sigma),t\bigr)$ for $\sigma \in S^2$, $t \in S^1$
where $\theta:S^1 \cong SO(2) \to \Diff(S^2)$ is the  group homomorphism
corresponding to the rotation of $S^2$ in $\bR^3$ around a suitably chosen rotation axis $\hat{a} \in \bR^3$
(and where we have set $\theta_t:=\theta(t)$ for $t \in S^1$).
Let $\Check{L}$ be the link obtained from $\hat{L}$ by the deformation $\phi$.
\begin{figure}[h]
\begin{center}
\includegraphics[height=1.5in,width=2.5in]{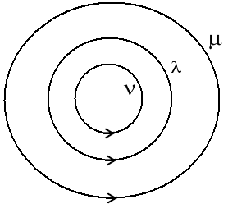}
 \caption{} \label{3concentric}
\end{center}
\end{figure}
For a suitable choice of the rotation axis $\hat{a}$  the
projection of $\Check{L}$  onto  $\Sigma=S^2$ will look like in Fig. \ref{3concentric}.
From the invariance properties of the
 Chern-Simons path integral we conclude at a heuristic level that
\begin{equation}\WLO(\Check{L}) = \WLO(\hat{L}) = \frac{1}{|X_{\emptyset}|}
\frac{1}{S_{00}^2} N_{\l\m\n}
\end{equation}
On the other hand,
after performing a suitable change of framing\footnote{alternatively,
we can replace the two changes of framing by
two  simple surgery operations. The first surgery operation involves a
suitable tubular neighborhood  of the
vertical loop $l_2$ in  Fig. \ref{smallcircles2} (cf. Sec. 4.2 in \cite{Wi});
 the second surgery operations involves
a similar  tubular neighborhood of the vertical loop $l_3$}
 for each of the two loops
$l_2$ and $l_3$ (i.e. the two vertical loops
 with colors ``$\mu$'' and ``$\nu$'' in Fig. \ref{smallcircles2})
  the link $L$ becomes isotopic to  $\Check{L}$.
 According  to Eq. (2.33) in \cite{Wi}
(and the paragraph after Eq. (4.40)
in Sec. 4.5 in \cite{Wi} where the conformal weight $h$ appearing in (2.33) in \cite{Wi}
is related to the matrix $T$)  each of these two
changes of framing\footnote{similarly, according to  Sec. 4.5 in \cite{Wi}
 the two surgery operations mentioned in the previous footnote will alter the  value of the WLO
 by the factor $T_{\mu\mu} T_{\nu \nu}^{-1}$, so also by using
 the surgery argument we arrive at Eq. \eqref{eq_Witten}}
   alters the value of the WLO by a factor $\frac{T_{\mu
\mu}}{T_{00}}$ and $\frac{T_{\nu \nu}}{T_{00}}^{-1}$, i.e. we  have
\begin{equation} \label{eq_Witten} \WLO(L) = \frac{T_{\mu \mu}}{T_{\nu \nu}}  \WLO(\Check{L}) =
 \frac{1}{|X_{\emptyset}|}  \frac{1}{S_{00}^2}
 \frac{T_{\mu \mu}}{T_{\nu \nu}} N_{\l\m\n}
\end{equation}

\noindent {\bf Conclusion:} By combining the two equations
\eqref{eq_our} and \eqref{eq_Witten} we obtain
\begin{equation} \label{eq_our'}
{T_{\m\m}\over T_{\n\n}}\, N_{\l\m\n} =\sum_{\lambda_0,\lambda_1}
 M_{ \lambda  \lambda_0}^{\lambda_1} \
S_{\nu \lambda_0} S_{\mu \lambda_1}  T_{\lambda_0 \lambda_0}
T_{\lambda_1 \lambda_1}^{-1}
\end{equation}

On the other hand according to Eq. \eqref{eq_2.step} in Example
\ref{Ex2} above we have
\begin{equation} \label{eq_not_our}
{T_{\m\m}\over T_{\n\n}}\, N_{\l\m\n} = \sum_{\lambda_0,\lambda_1}
 N_{ \lambda  \lambda_0}^{\lambda_1} \
S_{\nu \lambda_0} S_{\mu \lambda_1}  T_{\lambda_0 \lambda_0}
T_{\lambda_1 \lambda_1}^{-1}
\end{equation}
 Eq. \eqref{eq_our'} and  Eq. \eqref{eq_not_our}  imply $\sum_{\lambda_0,\lambda_1}
 M_{ \lambda  \lambda_0}^{\lambda_1} \
S_{\nu \lambda_0} S_{\mu \lambda_1}  T_{\lambda_0 \lambda_0}
T_{\lambda_1 \lambda_1}^{-1} = \sum_{\lambda_0,\lambda_1}
 N_{ \lambda  \lambda_0}^{\lambda_1} \
S_{\nu \lambda_0} S_{\mu \lambda_1}  T_{\lambda_0 \lambda_0}
T_{\lambda_1 \lambda_1}^{-1}$. This holds for  arbitrary $\l,
\m, \n \in \Lambda^k_+$  so using the fact that S-matrix and the
T-matrix are invertible (cf. Eqs. \eqref{defST_0} above) we indeed
obtain $M^{\l}_{\m\n} =N^{\l}_{\m\n}$ (for all $\l, \m, \n \in
\Lambda^k_+$).

\begin{remark} \rm \label{rem6}
 Witten's argument from Sec 4.5 in \cite{Wi} which we used in
the paragraph preceding Eq. \eqref{eq_Witten} is based on ideas
from conformal field theory. So if we want to give a (complete)
path integral derivation of Eq. \eqref{eq_quantum_racah} we will
have to derive the first equality in Eq. \eqref{eq_Witten} using
only path integral methods, cf. \cite{HaSe} for partial results
in this direction.  On the other hand,
 if one is happy with ``mixing''  arguments from conformal field
theory and arguments based on the CS path integral then the
derivation of the (elementary) quantum Racah formula Eq.
\eqref{eq_quantum_racah} which we have just given is fine and by
combining Eq. \eqref{eq_quantum_racah} with the Verlinde formula
derived in Remark \ref{rm_BlTh} (using  Eq. (4.36) in \cite{Wi})
one finally obtains $\hat{N}_{ \gamma \alpha}^{\beta} = \sum_{\tau
\in \cW_{k}} \sgn(\tau) m_{\gamma}(\alpha-\tau(\beta))$,
which is the affine Lie algebra version
of the  ``abstract''  quantum
Racah formula appearing at the end of Subsec. \ref{subsec2.2}.
\end{remark}

\section{Outlook}
\label{sec6}

In the introduction we mentioned one of the most
important open questions in the theory of 3-manifold quantum
invariants,  the question  whether and how one can make rigorous
sense of Witten's heuristic  path integral expressions for
the Wilson loop observables of Chern-Simons theory, cf.  the r.h.s. of Eq. \eqref{eq_WLO}.
 A related and probably  less difficult question is
  whether and how one can make rigorous sense of those path
integral expressions  that arise from the r.h.s. of Eq.
\eqref{eq_WLO} after choosing a suitable gauge fixing.
 Until recently Lorentz gauge fixing was the only\footnote{for CS models on the
 special manifold $M=\bR^3$ there is an alternative approach
 based on light-cone gauge fixing and a suitable complexification of
 the manifold, cf.  \cite{FK},
 \cite{Ka3}. However, in this approach certain correction factors
 have to be inserted ``by hand'' in the course of the
 computations. At present the origin of these correction factors is not  clear} gauge fixing procedure
 for which  the relevant path integral expressions have been evaluated completely for general
 groups, links and manifolds, cf.
 \cite{GMM,Bar,Bar2,AxSi1,BoTa,AxSi2,AlFr}.
The final result of this evaluation is a  complicated
 infinite  series whose
 terms involve integrals over  (high-dimensional)
 ``configuration spaces'', cf.  \cite{BoTa,AlFr}.
The heuristic path integral expressions which appear during the
intermediate computations are even more complicated and it should
be  hard to find a rigorous realization of these path
integrals\footnote{There are, however, some very interesting partial
results in this direction, cf. \cite{AlbMit}}.\par

It is therefore desirable to find other gauges for which the WLOs
can also be evaluated explicitly. A gauge which leads to the
expressions appearing in Turaev's shadow world approach to the
3-manifold quantum invariants  would be particularly desirable.
This is  because the expressions appearing in the shadow world
approach  involve only finite sums, which are defined in a purely
combinatorial
 way. These finite combinatorial sums  are considerably less complicated than the
 infinite series of configuration space integrals mentioned above.
 Accordingly, it is  reasonable to believe that for such a gauge fixing
also the corresponding path integral expressions and the heuristic
arguments used for their evaluation will be less complicated than
those for Lorentz gauge fixing.\par

 The results in \cite{Ha4} and the present paper suggest that for manifolds $M$
of the form $M=\Sigma \times S^1$ torus gauge fixing is a  gauge
fixing with the desired properties.
Moreover, it is reasonable to expect that the  path integral expressions for the WLOs in
the torus gauge, i.e. the r.h.s. of Eq. \eqref{eq_WLO_0} above,
 admit   a rigorous treatment, either in a ``continuum setting''
(cf.  Secs. 8--9 in  \cite{Ha3b}, Sec. 4--6 in  \cite{Ha4}, and \cite{Ha6}
for partial results) or in a suitable ``simplicial setting''
(cf. \cite{Ha7a,Ha7b} for ongoing work in this direction).

\bigskip

{\em Acknowledgements:} A. H. would like to thank Ambar N.
Sengupta and Stephen Sawin for useful discussions and the
Alexander von Humboldt foundation for the generous financial
support during the period Oct. 2005 -- Sept. 2006.

\bigskip

\begin{flushleft}
 {Amsterdam University College, University of Amsterdam\\
Plantage Muidergracht 14,  1018 TV Amsterdam}

 \smallskip

 and

\smallskip

Institute for Theoretical Physics, University of Amsterdam \\
Science Park 904, Postbus 94485,  1090 GL Amsterdam,  The Netherlands\\
{\em E-mail address:} s.deharo@auc.nl
\end{flushleft}

\medskip

\begin{flushleft}
Grupo de F{\'i}sica Matem{\'a}tica da Universidade de Lisboa\\
Av. Prof. Gama Pinto, 2\\
PT-1649-003 Lisbon,  Portugal\\
{\em E-mail address:}  atle.hahn@gmx.de
 \end{flushleft}


\begin{thebibliography}{99}
\addcontentsline{toc}{section}{References}

\bibitem{AOSV}
  M.~Aganagic, H.~Ooguri, N.~Saulina and C.~Vafa,
   ``Black holes, q-deformed 2d Yang-Mills, and non-perturbative topological
  strings,''
  Nucl.\ Phys.\ B { 715} (2005) 304
  [arXiv:hep-th/0411280].



\bibitem{ASen}
S.~Albeverio and A.N. Sengupta.
\newblock {A Mathematical Construction of the Non-Abelian Chern-Simons
  Functional Integral}.
\newblock {\em Commun. Math. Phys.}, 186:563--579, 1997.

\bibitem{AlbMit}
S.~Albeverio and I. Mitoma.
\newblock {Asymptotic expansion of perturbative Chern-Simons theory via Wiener space}.
{\em  Bull. Sci. Math.}, 133(3): 272–-314, 2009.


\bibitem{AlFr}
D.~Altschuler and L.~Freidel.
\newblock {Vassiliev knot invariants and Chern-Simons perturbation theory to
  all orders}.
\newblock {\em Comm. Math. Phys.}, 187:261--287, 1997.

\bibitem{And1} H. H. Andersen.
\newblock {Tensor products of quantized tilting modules}.
\newblock {\em Commun. Math. Phys.}, 149:149--159, 1992.

\bibitem{AndPar} H. H. Andersen. and J. Paradowski.
\newblock {Fusion categories arising from semisimple Lie algebras}.
\newblock {\em Comm. Math. Phys.}, 169(3):563--588, 1995.




\bibitem{AxSi1}
S.~Axelrod and I.M. Singer.
\newblock {Chern-Simons perturbation theory.}
\newblock In {Catto, Sultan et al.}, editor, {\em {Differential geometric
  methods in theoretical physics. Proceedings of the 20th international
  conference, June 3-7, 1991, New York City, NY, USA}}, volume 1-2, pages
  3--45. World Scientific, Singapore, 1992.


\bibitem{AxSi2}
S.~Axelrod and I.M. Singer.
\newblock {Chern-Simons perturbation theory. II.}
\newblock {\em J. Differ. Geom.}, 39(1):173--213, 1994.


\bibitem{Bar}
D.~Bar-Natan.
\newblock {Perturbative Chern-Simons theory}.
\newblock {\em J. Knot Th. Ram.}, 4:503--547, 1995.


\bibitem{Bar2}
D.~Bar-Natan.
\newblock {On the Vassiliev knot invariants}.
\newblock {\em Topology}, 34:423--472, 1995.

\bibitem{BW}
  C.~Beasley and E.~Witten,
\newblock{ Non-abelian localization for Chern-Simons theory}.
 {\em J. Differential Geom.}, 70(2): 183-–323, 2005. [arXiv:hep-th/0503126]

\bibitem{BlTh1}
M.~Blau and G.~Thompson.
\newblock {Derivation of the Verlinde Formula from Chern-Simons Theory and the
  G/G model}.
\newblock {\em Nucl. Phys.}, B408(1):345--390, 1993.

\bibitem{BlTh2}
M.~Blau and G.~Thompson.
\newblock {Lectures on 2d Gauge Theories: Topological Aspects and Path Integral
  Techniques}.
\newblock In E.~Gava et~al., editor, {\em Proceedings of the 1993 Trieste
  Summer School on High Energy Physics and Cosmology}, pages 175--244. World
  Scientific, Singapore, 1994.

\bibitem{BlTh3}
M.~Blau and G.~Thompson.
\newblock {On Diagonalization in $Map(M,G)$}.
\newblock {\em Commun. Math. Phys.}, 171:639--660, 1995.


\bibitem{BlTh4}
M.~Blau and G.~Thompson.
\newblock {Chern-Simons Theory on $S^1$-Bundles: Abelianisation and q-deformed Yang-Mills Theory},
{\em  J. High Energy Phys.}, no. 5, 003, 35 pp, 2006. [arXiv:math-ph/0601068]

\bibitem{BoTa}
R.~Bott and C.~Taubes.
\newblock {On the self-linking of knots}.
\newblock {\em J. Math. Phys.}, 35(10):5247--5287, 1994.

\bibitem{Br_tD}
Th. Br{\"o}cker and T.~tom Dieck.
\newblock {\em Representations of compact {L}ie groups}, volume~98 of {\em
  Graduate Texts in Mathematics}.
\newblock Springer-Verlag, New York, 1985.


\bibitem{BR}   E.~Buffenoir and P.~Roche,
  ``Two-dimensional lattice gauge theory based on a quantum group,''
  Commun.\ Math.\ Phys.\  { 170} (1995) 669
  [arXiv:hep-th/9405126].


\bibitem{deHa2}
S.~de~Haro.
\newblock {Chern-Simons theory on lens spaces from 2d Yang-Mills on the
  cylinder}.
\newblock {\em J. High. Energy Phys.}, 41(8), 2004.

\bibitem{deHa1}
S.~de~Haro.
\newblock{A Note on Knot Invariants and $q$-Deformed $2d$ Yang-Mills}.
\newblock{\em Phys. Lett. B}, 634 (2006) 78-83 [arXiv:hep-th/0509167]

\bibitem{deHa3}
  S.~de Haro,
  ``Chern-Simons theory, 2d Yang-Mills, and Lie algebra wanderers,''
  Nucl.\ Phys.\ B { 730} (2005) 312  [arXiv:hep-th/0412110].


\bibitem{deHaTi}
  S.~de Haro and M.~Tierz,
  ``Discrete and oscillatory matrix models in Chern-Simons theory,''
  Nucl.\ Phys.\ B { 731} (2005) 225
  [arXiv:hep-th/0501123].



\bibitem{diF} F.~Di Francesco, P.~Mathieu and D.~S\'en\'ecal, ``Conformal Field Theory'', Springer-Verlag, 1997


\bibitem{Finkelberg} M. Finkelberg.
\newblock {An equivalence of fusion categories}.
\newblock {\em Geom. Funct. Anal.}, 6:249--267, 1996.

\bibitem{Freed}
D.~S. Freed.
\newblock Quantum groups from path integrals. Preprint, 1995
\newblock [arXiv:q-alg/9501025]

\bibitem{FK}
J.~Fr{\"o}hlich and C.~King.
\newblock {The Chern-Simons Theory and Knot Polynomials}.
\newblock {\em Commun. Math. Phys.}, 126:167--199, 1989.


\bibitem{Fuchs} J. Fuchs and C. Schweigert.
\newblock A representation theoretic approach to the WZW Verlinde formula. Preprint, 1997
[arXiv:hep-th/9707069]


\bibitem{FuchsDriel} J. Fuchs, and P. van Driel.
 \newblock WZW fusion rules, quantum groups, and the modular matrix  S.
\newblock Nucl. Phys. B, 346:632--648, 1990.



\bibitem{GMM}
E.~Guadagnini, M.~Martellini, and M.~Mintchev.
\newblock {Wilson Lines in Chern-Simons theory and Link invariants}.
\newblock {\em Nucl. Phys. B}, 330:575--607, 1990.

\bibitem{Ha2}
A.~Hahn.
\newblock {The Wilson loop observables of Chern-Simons theory on ${\mathbb
  R}^3$ in axial gauge}.
\newblock {\em {Commun. Math. Phys.}}, 248(3):467--499, 2004.

\bibitem{Ha3b}
A.~Hahn.
\newblock {Chern-Simons models on $S^2 \times S^1$, torus gauge fixing, and
  link invariants I}.
\newblock {\em {J. Geom. Phys.}}, 53(3):275--314, 2005.


\bibitem{Ha3c} A.~Hahn.
\newblock {Chern-Simons models on $S^2 \times S^1$, torus gauge fixing, and
  link invariants II}.
\newblock {\em {J. Geom. Phys.}}, 58:1124--1136, 2008.


\bibitem{Ha4}
A.~Hahn.
\newblock {An analytic Approach to Turaev's Shadow Invariant}.
\newblock {\em J. Knot Th. Ram.}, 17(11): 1327--1385, 2008
[see arXiv:math-ph/0507040v7  for the most recent version]



\bibitem{Ha6}
A.~Hahn.
\newblock {White noise analysis in the theory of three-manifold quantum
  invariants}.
\newblock In A.N. Sengupta and P.~Sundar, editors, {\em Infinite Dimensional
  Stochastic Analysis}, volume XXII of {\em Quantum Probability and White Noise
  Analysis}, pages 201--225. World Scientific, 2008.


\bibitem{Ha7a}
A.~Hahn.
\newblock {From simplicial Chern-Simons theory to the shadow invariant I},
Preprint, 2012 [arXiv:1206.0439]

\bibitem{Ha7b}
A.~Hahn.
\newblock {From simplicial Chern-Simons theory to the shadow invariant II},
 Preprint, 2012 [arXiv:1206.0441]

\bibitem{HaSe}
A.~Hahn
\newblock {Surgery operations on the Chern-Simons path integral via conditional expectations,
} in preparation.


\bibitem{HKPS}
T.~Hida, H.-H. Kuo, J.~Potthoff, and L.~Streit.
\newblock {\em {White Noise. An infinite dimensional Calculus}}.
\newblock Dordrecht: Kluwer, 1993.


\bibitem{Ka}
L.~Kauffman.
\newblock {\em {Knots}}.
\newblock Singapore: World Scientific, 1993.


\bibitem{Ka3}
L.~Kauffman.
\newblock {Functional integration, Kontsevich integral and formal integration.}
\newblock {\em J. Korean Math. Soc.}, 38(2):437--468, 2001.



\bibitem{ReKi} A. N. Kirillov and N.Y. Reshetikhin.
Representations of the algebra $U_q(sl_2)$,
 $q$-orthogonal polynomials and invariants of links.
In V.G. Kac et al., editor, {\em Infinite Dimensional Lie Algebras and Groups},
Vol. 7 of {\em Advanced Ser. in Math. Phys.},   pages 285--339, 1988.


\bibitem{Kup}
G.~Kuperberg.
\newblock {Quantum invariants of knots and 3-manifolds (book review)}.
\newblock {\em Bull. Amer. Math. Soc.}, 33(1):107--110, 1996.


\bibitem{PoRe}
M.~Polyak and N.~Reshetikhin.
\newblock {On 2D Yang-Mills Theory and Invariants of Links}.
\newblock In Sternheimer~D. et~al, editor, {\em Deformation Theory and
  Symplectic Geometry}, pages 223--246. Kluwer Academic Publishers, 1997.

\bibitem{ReTu2}
N.Y. Reshetikhin and V.G. Turaev.
\newblock {Ribbon graphs and their invariants derived from quantum groups.}
\newblock {\em Commun. Math. Phys.}, 127:1--26, 1990.

\bibitem{ReTu1}
N.Y. Reshetikhin and V.G. Turaev.
\newblock {Invariants of three manifolds via link polynomials and quantum
  groups.}
\newblock {\em Invent. Math.}, 103:547--597, 1991.


\bibitem{Roz}
L. Rozansky.
\newblock {A contribution of the trivial connection to Jones polynomial
and Witten's invariant of 3d manifolds. I},
{\em Commun.Math.Phys.} 175:275--296,  1996. [arXiv:hep-th/9401061]

\bibitem{Sawin99}
S.~Sawin.
\newblock {Jones-Witten invariants for non-simply connected Lie groups and the geometry of the Weyl alcove}.
{\em Adv. Math.}, 165(1):1–34, 2002. [arXiv: math.QA/9905010]

\bibitem{Sawin03}
S.~Sawin.
\newblock {Quantum groups at roots of unity and modularity}.
{J. Knot Th. Ram.}, 15(10):1245–-1277, 2006. [arXiv:math.QA/0308281]

\bibitem{Sawin05}
S.~Sawin.
\newblock {Closed subsets of the Weyl alcove and TQFTs}.
{\em Pacific Journal of Mathematics}, 238  305-324, 2006. [arXiv:math.QA/0503692]

\bibitem{Sawon}
J.~Sawon.
\newblock { Perturbative expansion of Chern-Simons theory}.
\newblock {\em Geom. Topol. Monogr.} 8:145--166, 2006 [arXiv:math/0504495]


\bibitem{Spieg} M. Spiegelglas.
 \newblock Spin sums, fusion rules and correlators by filling.
\newblock Phys. Lett. B,  247:36--40, 1990.





\bibitem{turaevbook} V.~G.~Turaev, ``Quantum Invariants of Knots and 3-Manifolds'', 1994, De Gruyter

\bibitem{Tu2} V.~G.~Turaev, ``Shadow Links and Face Models of Statistical Mechanics'', J.~Differential
Geometry 36 (1992) 35-74




\bibitem{TuVi}
V.G. Turaev and O.~G. Viro.
\newblock {State sum invariants of 3-manifolds and quantum 6j-symbols}.
\newblock {\em Topology}, 31(4):865--902, 1992.


\bibitem{TuWe}
V.G. Turaev and H. Wenzel.
\newblock {Quantum invariants of 3-manifolds associated with classical simple Lie algebras}.
\newblock {\em Internat. J. Math}, 4:323--358, 1993.




\bibitem{Wal} M. Walton.
 \newblock Fusion rules in Wess-Zumino-Witten models
\newblock Nucl. Phys. B, 340:777--790, 1990.

\bibitem{Walk} K. Walker
 \newblock On Witten's 3-manifold invariants,
\newblock preprint, 1991


\bibitem{Wi}
E.~Witten.
\newblock {Quantum Field Theory and the Jones Polynomial}.
\newblock {\em Commun. Math. Phys.}, 121:351--399, 1989.


\end{thebibliography}
\end{document}